\documentclass[onecolumn]{emulateapj}

\usepackage{latexsym}
\usepackage{amssymb, amsthm, amsmath, amsfonts}
\usepackage{MnSymbol}
\usepackage{graphicx}
\usepackage{natbib}
\usepackage{float}
\usepackage{microtype}  
\usepackage{booktabs}  

\newcommand{\Msun}{\ensuremath{\,\text{M}_\odot}}
\newcommand{\degrees}{\ensuremath{^{\circ}}}
\newcommand{\chis}{\ensuremath{\chi^{2}}}
\newcommand{\dchis}{\ensuremath{\Delta\chi^{2}}}

\newcommand{\sigstat}{\ensuremath{\sigma}}
\newcommand{\sigdegen}{\ensuremath{\sigma_\text{degen}}}
\newcommand{\be}{\begin{equation}}
\newcommand{\ee}{\end{equation}}
\newcommand{\bd}{\begin{displaymath}} 
\newcommand{\ed}{\end{displaymath}}   

\slugcomment{Accepted by ApJ Oct 23, 2016}
\shorttitle{Mass-Radius Constraints}
\shortauthors{Stevens et al.}

\begin{document}

\title{Neutron Star Mass-Radius Constraints using Evolutionary Optimization}

\author{A.~L. Stevens\altaffilmark{1,2}, J.~D. Fiege\altaffilmark{3}, D.~A. Leahy\altaffilmark{4}, and S.~M. Morsink\altaffilmark{1}}

\altaffiltext{1}{Department of Physics, University of Alberta}
\altaffiltext{2}{Anton Pannekoek Institute for Astronomy, University of Amsterdam}
\altaffiltext{3}{Department of Physics and Astronomy, University of Manitoba}
\altaffiltext{4}{Department of Physics, University of Calgary}

\begin{abstract}

The equation of state of cold supra-nuclear-density matter, such as in neutron stars, is an open question in astrophysics. 
A promising method for constraining the neutron star equation of state is modelling pulse profiles of thermonuclear X-ray burst oscillations from hotspots on accreting neutron stars. 
The pulse profiles, constructed using spherical and oblate neutron star models, are comparable to what would be observed by a next-generation X-ray timing instrument like ASTROSAT, NICER, or LOFT. 
In this paper, we showcase the use of an evolutionary optimization algorithm to fit pulse profiles to determine the best-fit masses and radii. 
By fitting synthetic data, we assess how well the optimization algorithm can recover the input parameters. 
Multiple Poisson realizations of the synthetic pulse profiles, {  constructed with 1.6 million counts and no background}, were fitted with the Ferret algorithm to analyze both statistical and degeneracy-related uncertainty, and to explore how the goodness-of-fit depends on the input parameters.
{  For the regions of parameter space sampled by our tests,} the best-determined parameter is the projected velocity of the spot along the observer's line-of-sight, with an accuracy of $\le3$\% compared to the true value and with $\le5$\% {  statistical uncertainty}.
The next best-determined are the mass and radius; for a neutron star with a spin frequency of 600\,Hz, the best-fit mass and radius are accurate to $\le5$\%, with respective {  uncertainties} of $\le7$\% and $\le10$\%.
The accuracy and precision depend on the observer inclination and spot co-latitude, with values of $\sim1$\% achievable in mass and radius if both the inclination and co-latitude are $\gtrsim60\degrees$.

\end{abstract}


\section{Introduction} \label{s:intro}

Neutron stars are an astrophysical laboratory for studying cold supra-nuclear-density matter.
Accreting millisecond X-ray pulsars, a particular subset of neutron stars in low-mass X-ray binaries (LMXBs), are rapidly spinning accretion-powered neutron stars with spin periods of a few milliseconds (e.g., SAX J1808.4$-$3658, \citealt{WijnandsVanDerKlis98}).  
Their pulsed X-ray emission originates from material striking the surface of the neutron star during regular accretion and warming an area on the surface so that it emits blackbody radiation. 
Then, as the neutron star rotates, it gives periodic oscillations in brightness as the emitting region faces towards and away from the observer. 
Since these photons originate from the surface of the neutron star itself, physical properties like its mass and radius are encoded in the detected pulse profile. 
Fitting these pulse profiles with realistic models can then yield neutron star mass and radius estimates \citep{Watts16}.

In addition to regular pulsed X-ray emission, some neutron stars in LMXBs exhibit thermonuclear (Type I) X-ray bursts  \citep{Watts12} . 
In a fraction of thermonuclear X-ray bursts, we observe brightness oscillations, where the frequency corresponds strongly with the spin frequency of the neutron star; these are referred to as thermonuclear burst oscillations. 
The pulse profile models that we discuss in this paper refer specifically to models of these burst oscillations.

The emission area on the surface of the neutron star is referred to as the hotspot or spot. 
Theories suggest two different surface hotspot models: 
one that ignites nuclear burning at one point and spreads across the whole neutron star, and another that ignites at one point and begins to spread but remains limited to a smaller area (referred to as a ``persistent hotspot")  (see \citealt{Watts12} and references therein).
The persistent hotspot on the surface of a rotating neutron star has been demonstrated to be an effective model for Type I X-ray bursts from the source 4U 1636$-$536 \citep{Artigueetal13}, and so we use a persistent spot model with no size variation over the course of the burst. 
The fixed spot model is used for convenience here; a changing spot model could be incorporated for observations that show evidence for such behaviour.

The spectral model depends on the physics of the spot production and includes both the energy and angular dependence of the emitted radiation.
In the case of rotation-powered X-ray pulsars \citep{Bogdanov}, a hydrogen atmosphere model is appropriate (e.g., \citealt{Heinke}). 
The hydrogen atmosphere model depends on the spot's temperature and the local surface gravity. 
Since the surface gravity only depends on the mass, radius, and spin of the star \citep{AlGendy14}, the local temperature is the only additional free parameter introduced by the spectral model.

For accretion-powered X-ray pulsars, an empirical model including a blackbody plus Comptonized photons has been used, motivated by spectral observations \citep{PoutanenGierlinski03,Leahy09,Leahyetal11,MorsinkLeahy11}. 
The Comptonization model includes photon power law indices as well as a parametrized fan-beaming model. 
These models required two free parameters, one for the energy dependence and the second for the angular dependence of the emitted radiation. 
One of the issues seen with fitting the data from the accretion-powered X-ray pulsars is that the extra degree of freedom in the radiation's angular dependence leads to extra degeneracies amongst the geometric parameters. 
The result is fairly large regions of parameter space allowed by the fits, which do not strongly constrain the neutron star's equation of state. 

Since thermonuclear X-ray burst oscillations can be well-reproduced with a simpler spectral model, we can fit models in a reduced but still physically motivated parameter space.
Fewer free parameters gives fewer degeneracies among the parameters, and therefore allows for better constraints on the neutron star's mass and radius.

In the setup of our models, we assume that the inner boundary of the accretion disk is the neutron star's co-rotation radius \citep{GhoshLamb79}. 
From this assumption, the accretion disk would block emission from a possible second antipodal hotspot (from an observer's perspective), so we only test models for one spot in the northern hemisphere of the neutron star. 
If the signature of an antipodal spot is detected, our code can be easily adapted to include a second hotspot.

We created synthetic data for a variety of different neutron star and spot parameters. 
Fitting multiple Poisson realizations for each synthetic pulse profile allows us to disentangle what uncertainty is due to random statistical fluctuations and what is due to inherent degeneracy between the parameters; 
understanding both is crucial for placing proper constraints on neutron star masses and radii. 
The pulse profile fitting was carried out with the Ferret optimization algorithm \citep{Fiege10} to determine the acceptable range of masses and radii.

In this paper we show that evolutionary optimization algorithms are a powerful method of fitting neutron star pulse profiles, 
and we test the effects of changing various input parameters on how well we can recover the true neutron star mass and radius.
In Section 2 we explain the details of constructing the pulse profiles, our parameter choices, and the Ferret  algorithm.
The results of the pulse profile fitting are described and examined in Section 3, and the conclusions are discussed in Section 4.

\section{Pulse Profile Models}

We construct the pulse profiles within the Schwarzschild + Doppler (S+D) approximation \citep{MillerLamb98,PoutanenGierlinski03} and the Oblate Schwarzschild (OS) approximation \citep{Morsinketal07}. 
In the S+D and OS approximations, the metric exterior to the rotating neutron star is approximated by the Schwarzschild metric as described by \citet{Pechenicketal83}, adding the appropriate Doppler boost factors arising from the rotation of the star. 
In the S+D approximation, the surface of the star is a sphere, while in the OS approximation the surface is an oblate spheroid. 
\citet{Cadeauetal07} compared the S+D and OS results with pulse profiles generated from the exact metric and showed that the OS approximation is a good approximation for stars spinning with frequencies above 300 Hz. 
However, we continue to use the S+D approximation in many of our models in order to further explore the effect of using the wrong shape on the fits. 
At spin frequencies higher than 600 Hz, it may be necessary to use higher order approximations that make use of the star's quadrupole moment \citep{Psaltis14}, however this level of approximation is not necessary for the stars studied in this paper.

Pulse profiles can be constructed once 8 geometric parameters and a spectral emissitivity model are specified. 
The 8 geometric parameters are the neutron star's mass $M$; equatorial radius $R$;
spin frequency $\nu_\text{spin}$; the observer's inclination angle $i$ (as measured from the spin axis); the hotspot's co-latitude $\theta$; the angular radius of the spot $\rho$; the distance to the star $d$; and a phase offset $\phi$.  
In practice, the star's spin frequency will always be known, so there are only 7 geometric parameters.
It is possible to add parameters describing a more complicated shape for the spot \citep{Poutanen09}, but in this paper we only consider the simplest spot models, which are circular spots with uniform temperature. 
Due to the approximately universal nature of a spinning neutron star's shape \citep{Morsinketal07,Baubock13}, inclusion of the star's oblate shape does not require any additional free parameters. 

Thermonuclear X-ray burst oscillations can be spectrally modelled as a single-temperature blackbody with limb-darkening~\citep{Artigueetal13}. 
Once a prescription for the angular dependence has been selected, such as the \citet{Chandrasekhar60} limb-darkening model {  (approximated by the Hopf function)}, the only free parameter is the hotspot's temperature. 
The reduced parameter space required by this spectral model results in less degeneracy with the geometric parameters. 
For this reason, the oscillations from Type I X-ray bursts will be a major target for {  large-area, high-time-resolution} X-ray telescopes like ASTROSAT~\citep{Singh}, the soon-to-be launched NICER mission~\citep{Arzoumanian}, and {  a future LOFT-like} mission~\citep{Ferocietal12}.

The geometric parameters $M, R, i,$ and $\theta$ have inherent degeneracies, so it is useful to refer to less-degenerate combinations of these parameters. 
In the spherical S+D approximation, $i$ and $\theta$ only appear in the formulae in the combinations 
$\sin i \sin\theta$ and $\cos i \cos\theta$. 
As a result, in all fits there is a simple degeneracy that allows $i$ and $\theta$ to be switched. 
Likewise, $M$ and $R$ are somewhat degenerate, so the dimensionless compactness ratio $M/R$ can be better-constrained. 
These parameter combinations factor into the approximate bolometric pulse amplitude $\hbox{Amp}$  \citep{Beloborodov02},
\be \label{eqn:amp}
\hbox{Amp} = \frac{ (1-2M/R) \sin i \sin \theta}{ 2M/R + (1-2M/R) \cos i \cos \theta},
\ee
and the dimensionless projected velocity of the spot $\beta$,
\be \label{eqn:beta}
\beta = \frac{2 \pi R \nu_\text{spin}}{\sqrt{1-2M/R}} \sin i \sin \theta,
\ee
in geometric units ($G=c=1$). 
Due to the reduced degeneracy, it is possible to fit for $\sin i \sin\theta$, $\cos i \cos\theta$, $M/R$, \hbox{Amp}, and $\beta$ better than the individual parameters. 
{  In our models, we consider an infinitesimally small spot for simplicity. 
However, the $i-\theta$ degeneracy can be partially broken for models with a large spot, in which case the spot would span a range in $\theta$ but not in $i$.}

Equation (\ref{eqn:amp}) is only an approximate relation for the bolometric pulse amplitude. In reality, the pulse amplitude depends both on the emitted energy spectrum and the
energy bands at which the observations are made. There is no simple formula for the dependence of the pulse amplitude on energy, but it can be computed numerically. 
Since the 
pulse amplitude \hbox{Amp} depends on the observed photon energy, observations of the burst oscillations in two or more energy bands can provide stronger constraints than suggested by equation (\ref{eqn:amp}); 
for this reason, two energy bands are used in this work.
{  Additional information could be extracted with more energy bands, but we were unable to accommodate more bands.}
The projected velocity $\beta$ controls the asymmetry in the pulse profile through the Doppler boosting effect, as well as the phase lags between the hard and soft energy bands.  
For higher values of $\beta$, the pulse profiles are more asymmetric in rise and fall times, showing the effects of higher harmonics. 

Previous work \citep{MorsinkLeahy11, ALSmscthesis,Loetal13,Baubock15} has shown that for smaller spots, it is sufficient to compute the pulse profile assuming a point-source spot instead of an extended region. 
By adopting the point-source approximation, we {  simplify the calculation and} do not make use of the angular radius parameter $\rho$, {  reducing} the number of free geometric parameters to 6. 
Furthermore, we normalize the synthetic pulse profiles to have a mean of 1 to remove the dependence on the distance $d$, giving five free geometric parameters for each model.

Our goal is to constrain the neutron star's mass and radius based on fitting models to the pulse profile, and determine how the shape of the pulse profile affects how well we can constrain its $M$ and $R$.
By fitting synthetic data, the input parameters are known, so we can analyze how well the fitting can recover the true $M$ and $R$.
In this paper we investigate two sources of error in the constraints: {  degeneracy-related uncertainty and statistical uncertainty. Although the two types of uncertainties are 
coupled we have introduced two measures of the uncertainties
\sigdegen\  and  \sigstat which are affected differently by the degeneracy and the statistics.} 
The {  degeneracy-related uncertainty \sigdegen\ }arises from parameter degeneracies in each fit, whereas the standard deviation \sigstat\ is examined by simulating multiple Poisson realizations of a model and determining the mean and standard deviation over all the fits.
We explore how \sigstat\ and \sigdegen\ are affected by the values of \hbox{Amp} and $\beta$ for each synthetic pulse profile. 
We also quote the accuracy of the fit in $M$ and $R$, comparing the mean best-fit value for each model with the true value.

\subsection{Properties of Test Models}
\label{s:props}

\begin{table*} 
\caption{Summary of Test Models \label{tab:Theoretical}}
\centering
\footnotesize
\begin{tabular}{llcc lllll ccccc}
\toprule
Name & Description & Shape & $\nu_\text{spin}$ & $M$ & $R$ & $i$ & $\theta$ & $\phi$ & $M/R$ & $\sin i \sin\theta$ & $\cos i \cos\theta$ & Amp & $\beta$ \\
 &  &  & Hz & $\Msun$ & km & deg. & deg. &  & & & & & \\
\midrule
A & Fiducial & Sphere & 600 & 1.6&  12&   60& 20&  0& 0.197& 0.296&  0.470&  0.373& 0.057 \\ 
AO & Oblate & Oblate & 600 & 1.6&  12&   60& 20&  0& 0.197& 0.296&  0.470&  0.373& 0.057 \\ 
$\theta_{37}$ & $\theta=37\degrees$ & Sphere & 600 & 1.6&  12&   60& 37&  0& 0.197& 0.521&  0.399&  0.720& 0.101 \\ 
$\theta_{37}$O & $\theta=37\degrees$ & Oblate &  600 & 1.6&  12&   60& 37&  0& 0.197& 0.521&  0.399&  0.720& 0.101 \\ 
$\theta_{60}$ & $\theta=60\degrees$ & Sphere & 600 & 1.6&  12&   60& 60&  0& 0.197& 0.750&  0.250&  1.305& 0.145 \\ 
$(M/R)_{hi}$ & High $M/R$ & Sphere & 600 & 1.68&  11&   60& 20.8&  0& 0.226& 0.308&  0.467&  0.350& 0.057 \\ 
$(M/R)_{lo}$ & Low $M/R$ &  Sphere & 600 & 1.35&  13&   60& 19.8& 0& 0.153& 0.293&  0.471&  0.423& 0.057 \\ 
$\beta_{hi}(M/R)_{hi}$ & High $\beta$, high $M/R$  & Sphere & 600 & 1.68&  11&   60& 64   & 0& 0.226& 0.778&  0.219&  1.235& 0.145 \\
$\nu_{400}$ & Low $\nu_\text{spin}$ & Sphere& 400 &  1.60&  12&   60& 20& 0 & 0.197& 0.296&  0.470&  0.373& 0.038 \\ 
\bottomrule
\end{tabular}
\vspace{12pt}
\end{table*}

We computed pulse profiles for a set of nine test models with the parameter values listed in Table~\ref{tab:Theoretical}.
The properties that were kept the same for all test models are the spot's temperature, the phase offset $\phi$, the observer's inclination
angle $i$ and the distance to the star $d$. 
The spot emission model is a 2\,keV blackbody (in the {  frame comoving with the neutron star's surface})  with a limb-darkening atmosphere, approximated by the Hopf function \citep{Chandrasekhar60}, appropriate for a Thompson-scattered atmosphere.  
{  Since for real data we would independently have the spot temperature {\it at infinity} instead of in the star's {  comoving} frame, this parameter should be allowed to vary within a narrow range (explored in Section \ref{s:incorrect}), where the appropriate range would be determined by the temperature at infinity.}
The model assumes that any emission from the surface of the neutron star outside the spot is negligible.
The observers inclination angle $i=60\degrees$ and the phase offset $\phi=0$ for all cases. 
The computed pulse profiles are normalized to an average flux of 1, so that the distance $d$ to the star does not affect the pulse profiles. 

The parameters that were changed for different test models are $M$, $R$, $\theta$, the star's shape (spherical or oblate), and $\nu_\text{spin}$.
For oblate neutron star models, $R$ is defined to be the radius of the neutron star at the spot. 
The formalism for the oblate model is detailed in \cite{Morsinketal07}, and does not require the addition of any extra parameters.

The parameters for the fiducial model, Model A, were chosen to be representative of the masses and radii of accreting millisecond neutron stars. 
Our fiducial mass $M=1.6\,\Msun$ is larger than the mass $M=1.4\,\Msun$ typically measured for slowly rotating radio pulsars, since we expect the neutron star has been spun up by and gained mass from accretion. 
The radius of 12 km is consistent with other radius estimations \citep{Leahy04,Steiner10}. 
Rapid rotation with $\nu_\text{spin}\geq 550\,$Hz is seen in burst oscillations for at least seven neutron stars that exhibit Type I X-ray bursts \citep{Watts12}.
We have chosen $\nu_\text{spin}=600\,$Hz for the fiducial model since it is representative of these rapid rotators. 
The angles were chosen to provide a pulse amplitude (Amp $=0.373$) comparable to the largest pulse amplitudes seen, in order to reduce the parameter degeneracy.
In particular, the neutron star 4U 1636-536 has rms pulse amplitudes as high as 0.25 {  after subtracting the pre-burst emission \citep{Strohmayeretal98, Galloway}}, where \hbox{Amp} $=\sqrt{2}\,\,\text{rms}$. 
The full set of theoretical parameters are given in the row labelled ``A" in Table \ref{tab:Theoretical}. 
The resulting pulse profiles in two energy bands are shown in Figure \ref{fig:A1data} with black and red solid curves labelled ``True". 
Our fiducial model is similar to the model described as ``low inclination" by \citet{Loetal13}, except that they consider a slightly smaller $R$ and a slower $\nu_\text{spin}$.
As will be shown in Section \ref{s:beta}, small changes in $R$ do not qualitatively change the results of this paper. 
Our choice of a faster $\nu_\text{spin}$ improves the accuracy of the fits, but the faster $\nu_\text{spin}$ is also more appropriate for the neutron stars that we hope to apply this method to.

Theoretical pulse profiles were calculated for the nine different test models.
Each theoretical pulse profile was converted to a set of twenty synthetic observations by adding noise from a Poisson distribution to the pulse profile. 
The standard case (a low-noise model) assumes 25,000 photon counts per phase bin {  and no background count rate, as if the background were negligible in comparison with the hot spot emission.
By not including a background, we are underestimating the Poisson fluctuations, and thus underestimating the error bars.
This is done so we can test the suitability of evolutionary optimization on the best possible quality of synthetic data. 

For a variation on the model $\theta_{60}$, a higher noise level was used: $\theta_{60}$C$_{6250}$ assumes 6250 photon counts per phase bin with no background. 
In total, there are ten sets of twenty simulated observed pulse shapes.

Each of the 200 simulated pulse profiles was then fit using the Ferret algorithm (described in the next subsection) which searches for different values of the free parameters in order to minimize the $\chis$ fit statistic. 
Five parameters were allowed to vary within physical ranges: $1.0 \Msun \le M \le 2.5 \Msun$, 6.0\,km $\le R \le 16.0$\,km, $0\degrees \le (i,\,\theta) \le 90\degrees$, 
and $\phi$ defined cyclically from 0 to $2\pi$.
The star's shape (spherical or oblate) was {  fixed to be the same} in the fitting procedure as in the theoretical waveform. 
The temperature is kept fixed at 2 keV since it has been shown by \citet{Loetal13} that observations in multiple energy bands allows for a good determination of the gravitationally-redshifted spot temperature. 
We do not allow the spot size to vary, as explained in the previous subsection.
For applications to real data, it would not be difficult to add variations in spot size (and shape), temperature, and to use unnormalized fluxes to allow for a distance measurement. 
However, the addition of the extra free parameters for the number of synthetic waveforms considered in this study would be impractical. 
We carried out preliminary trials (summarized in Section~\ref{s:incorrect}) allowing these parameters to vary but did not find much change in our final results.

Since there are two energy bands with 32 phase bins each, and five free parameters, there are 59 degrees of freedom (dof). 
The fitting yielded best-fit model parameters and also allowed computation of confidence regions in the $M$-$R$ plane, from which we can assess the effect of various parameters on the uncertainties in parameter determination.

\subsection{Parameter Fitting using Evolutionary Optimization}

Evolutionary optimization algorithms provide a useful but heuristic approach to search through large parameter spaces, to optimize the fit of a model to data.  
Such algorithms use principles inspired by biology to evolve a population of candidate parameters sets, over many generations, toward an optimal solution.  
The heuristic nature of the search aims to sample the parameter space efficiently, but not completely, since exhaustive search is not practical for problems involving many parameters.  
Therefore, it cannot be guaranteed that the true, globally optimal solution has been found, although this would also be true for any other optimization algorithm, besides exhaustive search.  
Evolutionary algorithms have been well studied and found to be useful in many fields of science and engineering. 

The oldest, and most commonly known, type of evolutionary optimizer is the genetic algorithm \citep{Holland, Goldberg89, Goldberg02}.  
Classic genetic algorithms encode their search parameters on a (usually) binary string (genotype), which is decoded into a model (phenotype).  
A population of candidate parameter sets is normally initialized as a set of random bit strings, which are expressed as a model, and in turn evaluated by a fitness function.  
This information is used to probabilistically select good parameter sets (individuals) to propagate to the next generation, in analogy to the ``survival of the fittest'' principle of Darwinian evolution.  
Parameter sets undergo random bitwise mutations to help perturb solutions into previously unexplored regions of parameter space.  
Information is shared between individuals by means of a crossover operator that cuts a pair of bit strings at a random position, and recombines them into a new ``offspring'' configuration, in analogy with sexual reproduction.  
This process of evaluation, selection, mutation, and selection is performed iteratively, over many generations, until a convergence criterion terminates the search.  
The genetic algorithm can be viewed as a directed stochastic search, which makes use of random noise to evade local minima, while a mostly deterministic selection operator pushes solutions toward an optimal solution.

In this paper, we used versions 5.3-5.5 of Ferret from the Qubist Global Optimization Toolbox for MATLAB \citep{Fiege10,RogersFiege11}{, a commercially available software package,} to find the best fits to our synthetic data.
{  This is} an alternative to the Markov chain Monte Carlo approach in Lo et al. (2013) to fit pulse profiles.  
Ferret's development began in 2002, as a variant of the multi-objective genetic algorithm, which made use of a real-valued parameter encoding and real-valued mutation and selection operators, rather than the traditional binary encoding discussed above.  
Numerous other features have been added to Ferret since then, which go beyond the usual genetic algorithm paradigm.  
Most notably, the code contains a unique linkage-learning algorithm, which detects a certain type of non-linearity (linkage) between parameters, with the goal of dividing a large parameter space into several smaller and nearly independent subspaces, thus greatly simplifying the problem.  
This capability is discussed at length in the software user's manual \citep{Fiege10}.  
Ferret also contains an algorithm that allows several of its most important control parameters, including the typical strength of mutation and crossover events, to be automatically adapted and optimized during a run.  
This auto-adaptation capability is similar in spirit to the self-adaptation used in Evolution Strategies (ES) codes, although ES codes do not typically employ a crossover operator.  
Ferret {  does not adhere to the strict definition of a genetic algorithm, due to these enhancements and others, but the code still remains closer to the genetic algorithm paradigm than to others within the family of evolutionary optimizers. }

{  As an example of Ferret's inner workings, consider the parameters $i$ and $\theta$. 
Ferret selects sets of values from the allowed region of parameter space ($0.01\degrees \leq i, \theta \leq 90\degrees$) for the population of a generation, and fits pulse profile models. 
It then keeps the best-fitting parameter sets and selects new sets to explore more of the parameter space. 
Within a few generations, Ferret discovers the degeneracy between $i$ and $\theta$, since their values can be swapped with minimal impact on the \chis\ fit.
A more detailed explanation of Ferret in relation to pulse profile modelling can be found in \citet{ALSmscthesis} Chapter 4.
Evolutionary optimization algorithms have also been used in gravitational lensing \citep{RogersFiege11, RogersFiege12}, medical physics \citep{Fiegeetal11}, star formation \citep{Franzmann14}, and X-ray spectral fitting \citep{Rogersetal15} applications.}

Our problem is quite easy for Ferret, which consistently finds the global minimum $\chi^2$ fit statistic within a few generations.  
We note that the true global minimum is known, since our results are based on artificial data tests.  
After finding the $\chi^2$ minimum, the algorithm was allowed to run for approximately 100 generations, as Ferret accumulated solutions within the confidence region.  
No specific convergence criterion was implemented; 
rather, we terminated the run manually when it was evident from the software's graphical user interface that no further improvements were being made to the minimum $\chi^2$, and enough solutions had been accumulated within the confidence region to make contours maps of acceptable quality.  
Ferret was executed in MATLAB R2011a on 12-core AMD Linux servers (dual socket Opteron 2439 SE, with 32Gb RAM, running Red Hat version 4.1.2), using Ferret's built-in parallel computing features.  
Each fit took approximately 8-20 hours.

\section{Results}

Pulse profiles were simulated as a set of  Poisson realizations of a theoretical pulse profile model with known input parameters (see Section~\ref{s:props}).
Each Poisson realization  was fit with Ferret to produce a set of best-fit parameters for the pulse profile along with confidence regions for the parameters. 
Since each theoretical pulse profile has multiple Poisson realizations, the average and standard deviation for the best-fit parameters of each theoretical pulse profile are computed. 

Here we discuss the fit results of the pulse profiles, which illustrate the effects of changing different system properties.
Table \ref{tab:summary} shows a summary of the fits to the 20 different Poisson realizations for each of the input models.
The first row of each pair of rows shows the input parameters of the theoretical pulse shape model. 
The second row shows the means and standard deviations of the best-fit values from the fits to the 20 different Poisson realizations of each model.
In addition to the fit parameters ($M$, $R$, $i$, $\theta$, and $\phi$),
other useful measures of the model ($M/R$, $\sin i \sin\theta$, $\cos i \cos\theta$, \hbox{Amp}, and $\beta$) are given. 

A brief overview of the entire set of fits is given first.
Some parameters are well-determined and others poorly-determined.
The standard deviation of any given parameter for each set of Poisson realizations is given as the error \sigstat\ of that parameter. 
We take the difference between the mean of a parameter for each set and the input parameter as a measure of the accuracy of that parameter fit. 
The means and standard deviations are listed in Table \ref{tab:summary}, from which the percent error for precision and accuracy can be determined. 
{  The Ferret algorithm also computes contour regions as a measure of the degeneracy-related uncertainty \sigdegen.
The \sigstat\ errors are sometimes smaller than the 1-\sigdegen\ region as shown in figures in the following subsections. 
These two values of error are measures of the degeneracy (\sigdegen) and the statistical fluctuations (\sigstat), but they need not be the same. 
The 1-\sigdegen\ contours determined by Ferret are models with different $i$, $\theta$, and $\phi$ parameters that all provide an equivalently good fit for that $M$ and $R$. 
In this way, the 1-\sigdegen\ limit can be thought of as a measure of the degeneracy of the parameter space near the best-fit model. 
The standard deviations \sigstat\ computed from the ensemble of Poisson realizations provide a measure of the error from the statistical fluctuations. 
However, the strong parameter degeneracy is certainly still an important factor in the standard deviation computation, and as such the reported \sigstat\ values cannot be assumed to be purely statistical error. }

We find that the projected velocity $\beta$ is generally the most accurate and precise, with accuracy of $0.5-3$\% and precision of $1-6$\%. 
$M$ and $R$ are the next best, with with accuracy of $1-8$\%, but $R$ has worse precision ($1-21$\%) than $M$ ($1-13$\%).
Compactness ($M/R$), $\sin i\sin\theta$,  and Amp all have similar accuracies ($\sim1-8$\%) and precisions ($\sim1-16$\%).
The accuracy of $\cos i\cos\theta$  is $\sim4-30$\% and precision is $\sim2-20$\%.
The least-well-determined are $i$ and $\theta$, with accuracies of $\sim4-30$\%  and precisions of $\sim5-50$\% due to the degeneracy in angles. 

In the following subsections we discuss in detail the results from fitting the fiducial model A, and then discuss trends we find by comparing the fits from model A with the fits for the models with modified parameters.

\begin{table*} 
\centering
\tiny
\caption{Summary of Statistical Properties of Fits \label{tab:summary}}
\begin{tabular}{lllllllllll}
\toprule
Model & $M$ & $R$ & $i$ & $\theta$ & $\phi$ & $M/R$ & $\sin i \sin\theta$ & $\cos i \cos\theta$ & \hbox{Amp} & $\beta$\\
\chis (59 dof) & \Msun & km & deg. & deg. & $\times 10^{-3}$ & & & & & $\times10^{-2}$ \\
\midrule[\heavyrulewidth]
A & 1.60&  12.0& 60.0& 20.0& 0.0 & 0.1969& 0.296&  0.470&  0.373& 5.74 \\ 
 59.3$\pm$7.6 & 1.52$\pm$0.11 &  11.4$\pm$1.2&   34.7$\pm$15.0& 41.8$\pm$15.2& 2.8$\pm$4.6 & 0.199$\pm$0.016& 0.318$\pm$0.032&  0.549$\pm$0.120&  0.363$\pm$0.025& 5.81$\pm$0.20 \\ 
\midrule  
AO & 1.60&  12.0& 60.0& 20.0& 0.0 & 0.1969& 0.296&  0.470&  0.373& 5.74 \\ 
 59.1$\pm$7.0 & 1.65$\pm$0.06 &  11.7$\pm$0.9&   49.3$\pm$18.2& 28.3$\pm$11.9& 0.8$\pm$5.3 & 0.209$\pm$0.016& 0.299$\pm$0.019&  0.511$\pm$0.142&  0.353$\pm$0.060& 5.76$\pm$0.18 \\ 
\midrule
$\theta_{37}$ & 1.60&  12.0&   60.0& 37.0& 0.0 & 0.1969& 0.521&  0.399&  0.720& 10.10 \\ 
 61.1$\pm$9.9 & 1.62$\pm$0.08 &  12.4$\pm$0.9&   49.1$\pm$13.8& 47.9$\pm$13.0& -0.7$\pm$2.9 & 0.194$\pm$0.009& 0.509$\pm$0.033&  0.386$\pm$0.064&  0.726$\pm$0.029& 10.09$\pm$0.17 \\ 
\midrule 
$\theta_{37}$O & 1.60&  12.0&   60.0& 37.0& 0.0 & 0.1969& 0.521&  0.399&  0.720& 10.10 \\ 
58.9$\pm$8.1 & 1.58$\pm$0.04 &  11.8$\pm$0.4&   49.8$\pm$9.7& 46.3$\pm$7.8& 1.1$\pm$1.5 & 0.198$\pm$0.008& 0.528$\pm$0.016&  0.424$\pm$0.029&  0.704$\pm$0.037& 10.08$\pm$0.14 \\ 
\midrule  
$\theta_{60}$ & 1.60&  12.0&   60.0& 60.0& 0.0 & 0.1969& 0.750&  0.250&  1.305& 14.54 \\ 
 57.7$\pm$8.4 & 1.61$\pm$0.01 &  12.1$\pm$0.1&   61.5$\pm$4.0& 58.5$\pm$3.7& -0.3$\pm$0.6 & 0.196$\pm$0.003& 0.745$\pm$0.008&  0.245$\pm$0.010&  1.309$\pm$0.011& 14.53$\pm$0.09 \\ 
\midrule
$\theta_{60}$C$_{6250}$ & 1.60&  12.0& 60.0& 60.0& 0.0 & 0.1969& 0.750&  0.250&  1.305& 14.54 \\ 
 58.2$\pm$7.7 & 1.62$\pm$0.03 &  12.5$\pm$0.5& 58.6$\pm$6.6& 60.2$\pm$6.6& -0.7$\pm$1.5 & 0.192$\pm$0.008& 0.728$\pm$0.027&  0.247$\pm$0.023&  1.303$\pm$0.028& 14.52$\pm$0.18 \\ 
\midrule  
$(M/R)_{hi}$ & 1.68&  11.0& 60.0& 20.8& 0.0 & 0.2255& 0.308&  0.467&  0.350& 5.74 \\ 
 60.3$\pm$7.7 & 1.61$\pm$0.11 &  10.4$\pm$1.0&   34.9$\pm$13.1& 42.7$\pm$16.2& 2.1$\pm$4.8 & 0.228$\pm$0.012& 0.328$\pm$0.029&  0.542$\pm$0.116&  0.343$\pm$0.022& 5.80$\pm$0.22 \\ 
\midrule 
$(M/R)_{lo}$ & 1.35&  13.0& 60.0& 19.8& 0.0 & 0.1534& 0.293&  0.471&  0.423& 5.75 \\ 
 62.5$\pm$10.1 & 1.28$\pm$0.10 &  13.0$\pm$1.2&   34.3$\pm$17.2& 42.8$\pm$17.1& 1.9$\pm$3.6 & 0.146$\pm$0.018& 0.299$\pm$0.027&  0.522$\pm$0.086&  0.411$\pm$0.022& 5.78$\pm$0.15 \\ 
\midrule 
$\beta_{hi}(M/R)_{hi}$ & 1.68& 11.0& 60.0& 64.0& 0.0 & 0.2255& 0.778&  0.219&  1.235& 14.53 \\ 
 53.8$\pm$10.7 & 1.70$\pm$0.02 &  11.2$\pm$0.3&   61.9$\pm$6.5& 61.6$\pm$6.8& -0.8$\pm$0.8 & 0.223$\pm$0.004& 0.762$\pm$0.016&  0.211$\pm$0.016&  1.241$\pm$0.016& 14.47$\pm$0.09 \\ 
\midrule  
$\nu_{400}$ & 1.60&  12.0&   60.0& 20.0& 0.0 & 0.1969& 0.296&  0.470&  0.373& 3.83 \\ 
58.1$\pm$9.9 & 1.55$\pm$0.21 &  13.0$\pm$2.5&   35.7$\pm$17.6& 40.7$\pm$18.8& 1.0$\pm$5.2 & 0.181$\pm$0.034& 0.292$\pm$0.051&  0.525$\pm$0.155&  0.361$\pm$0.035& 3.88$\pm$0.22 \\
\bottomrule
\end{tabular}
\vspace{12pt}
\end{table*}

\subsection{Model A: Fiducial Model}

\begin{figure}[tb]
\centering
\epsscale{1.0}
\includegraphics[width=0.49\columnwidth]{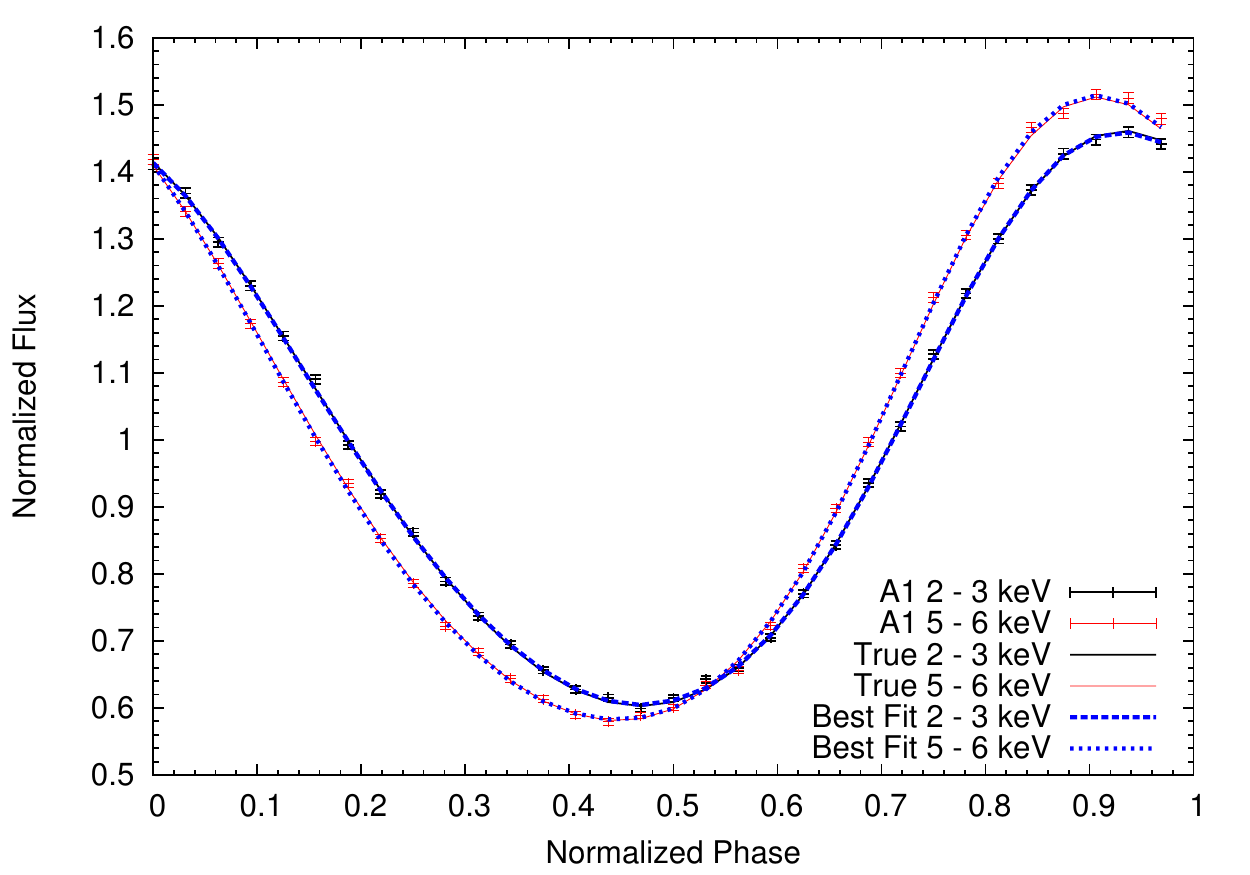}\includegraphics[trim=2cm 6.5cm 2cm 2cm, clip=true, width=0.45\columnwidth]{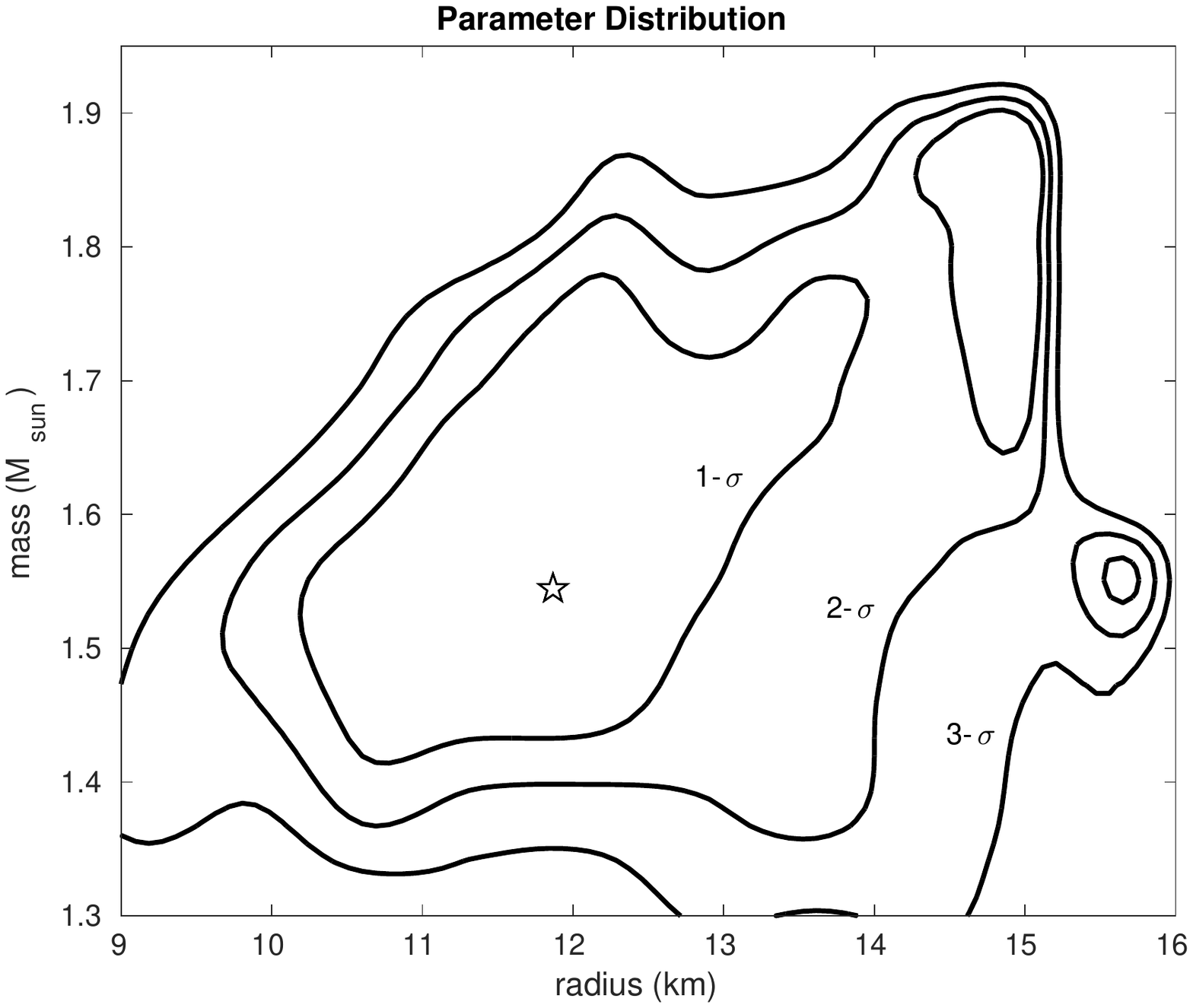}
\caption{\textsc{Left:} Pulse profiles for model A in two energy bands. 
The solid black and red curves (labeled ``True") correspond to the theoretical model A given in Table \ref{tab:Theoretical}. 
The black and red points (labeled ``A1") with error bars correspond to the addition of noise to model A with one Poisson realization assuming an average count-rate of 25,000 per phase bin.  
The blue dashed and dotted curves correspond to the best fit to model A1 with parameters given by the values in row ``A1" in Table \ref{tab:Adata}. 
This best fit has $\chis=58.2$ for 59 dof.
\textsc{Right:} \chis\ contours in the $M-R$ plane for fitting synthetic data from model A1. 
The best-fit model is shown with a star and corresponds to the row labelled ``1'' in Table \ref{tab:Adata}. The ``true" values of mass and radius are 1.6~$\Msun$ and 12~km.
The contours show the 1-, 2-, and 3-\sigdegen\ confidence regions.
From this figure, it can be seen that the 1-\sigdegen\ error region for radius corresponds to approximately 1.9 km, while the 1-\sigdegen\ limit for mass spans 0.18\Msun. 
The contours for 1-, 2-, and 3-\sigdegen\ correspond to values of \dchis\ (above $\chis_\text{min}$)  of 2.3, 6.2, and 11.8 respectively. 
}
\label{fig:A1data}
\end{figure}

\begin{figure}
\centering
\epsscale{1.0}
\includegraphics[width=0.59\columnwidth]{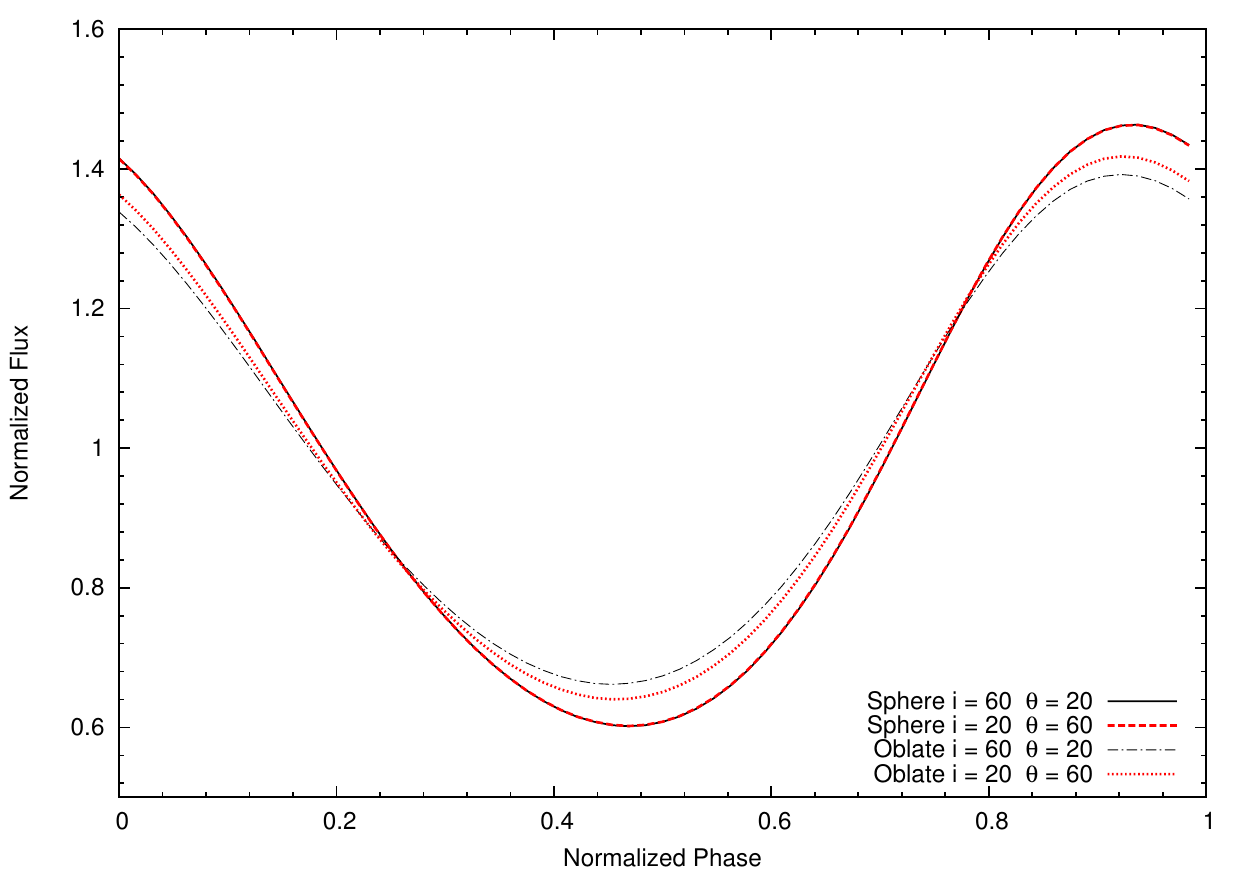}
\caption{
Normalized low-energy band pulse profiles for four neutron star models showing the
observer inclination--spot co-latitude degeneracy for spherical and oblate stars. 
The solid black curve represents Model A, while the overlapping dashed red curve represents a star with the same parameters as Model A, but with inclination $i$ and co-latitude $\theta$ swapped. 
The black dot-dashed curve represents Model AO, the oblate version of Model A. 
The red dotted curve represents a star with the same parameters as Model AO, again with inclination $i$ and co-latitude $\theta$ swapped.
}
\label{fig:degeneracy}
\end{figure}

\begin{figure}[tb]
\epsscale{1.0}
\centering
\includegraphics[width=0.47\columnwidth]{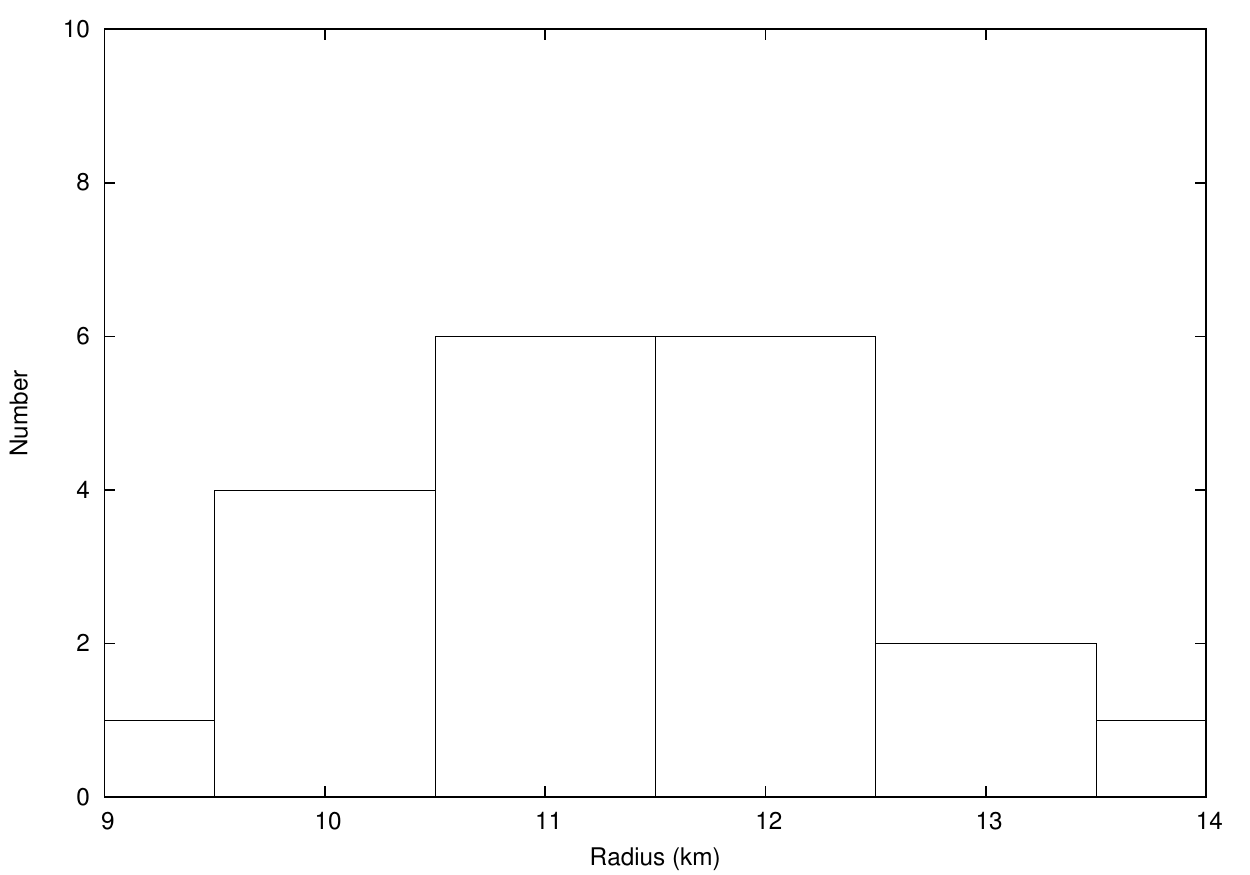} 
\includegraphics[width=0.47\columnwidth]{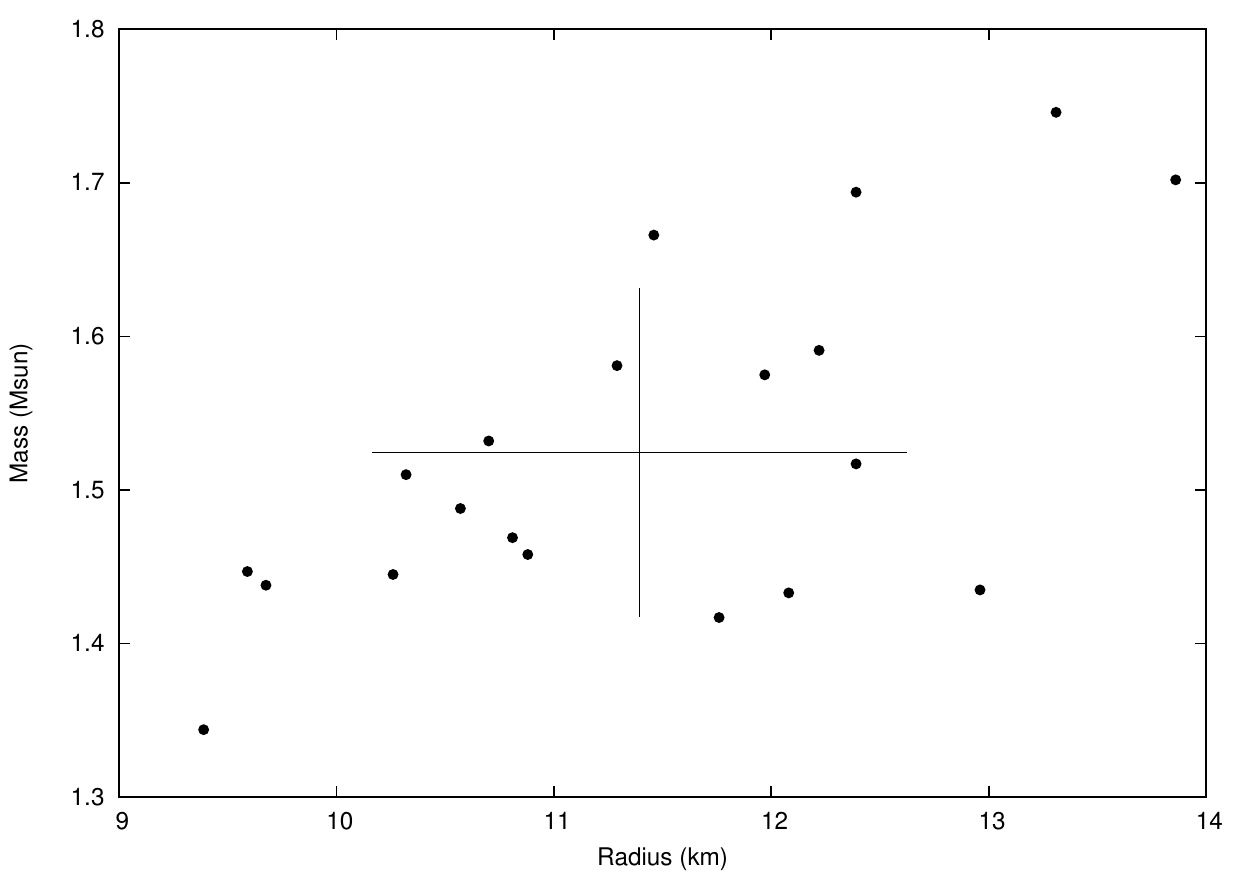}
\caption{\textsc{Left:} Histogram of best-fit values of $R$ for model A. 
\textsc{Right:} Best fit values of $M$ and $R$ for model A. 
Each dot is one fit to a Poisson realization, as shown in Table \ref{tab:Adata}. 
The black cross shows the average with 1-\sigstat\ error bars.
}
\label{fig:Ahistogram}
\end{figure}

Twenty realizations of the Poisson noise were made to simulate observations of the fiducial Model A. 
The resulting flux values with error bars for the first Poisson realization of the data are shown as black and red points labelled ``A1" on Figure \ref{fig:A1data}. 
The data points with Poisson errors were used as input to Ferret, which searched for the best fit to the A1 dataset by minimizing \chis. 
In the case of the A1 dataset, the pulse profiles that best fit the data are displayed with blue dashed and dotted curves in Figure~\ref{fig:A1data}. 
The parameters corresponding to the best fit of the A1 dataset ($\chis=58.2$) are shown in the row labelled ``1" in Table \ref{tab:Adata}. 
For reference we have included the derived quantities $M/R$, $\sin i \sin\theta$, $\cos i \cos\theta$, Amp, and $\beta$ to this table.  
 
The best fit model (star) and contours of constant $\chis = \chis_\text{min} + \dchis$ (black curves) for the A1 dataset are shown in the $M-R$ plane in Figure \ref{fig:A1data}. 
The contours correspond to values of $\dchis = 2.3, 6.2$, and $11.8$, corresponding to 1-, 2-, and 3-\sigdegen\ confidence levels respectively in mass-radius space for two free parameters, $M$ and $R$.
{  This plot uses the profile likelihood \citep{MurphyvdVaart} to eliminate the nuisance parameters $i$, $\theta$, and $\phi$; a grid is defined in $M$ and $R$, and for each $(M,R)$ grid point, Ferret finds the lowest \chis\ within each grid cell, allowing for any values of $i$, $\theta$, and $\phi$.}
The 1-\sigdegen\ confidence region (calculated by finding the maximum and minimum values on the 1-\sigdegen\ contour and dividing by 2) gives an approximate 1-\sigdegen\ error of 1.9\,km for the radius and a 0.18\Msun\ error for the mass. 
Note that since the contours are not ellipses, these are only approximate 1-\sigdegen\ limits.

Ferret was used to fit each of the 20 Poisson realizations of model A. 
The independent best fit results for the 20 different realizations are shown in Table \ref{tab:Adata}. 
The average and standard deviation for each parameter are displayed at the bottom of Table \ref{tab:Adata}, and also appear in the second line of Table \ref{tab:summary}. 
For the individual angles $i$ and $\theta$, it can be seen from Table \ref{tab:Adata} that the 
determinations of these angles are very poor. 
This is due to the already well-known degeneracy (for spherical stars; \citealt{PoutanenGierlinski03}) which occurs since the equations for light-bending and the Doppler effect only depend on the combinations $\sin i\sin\theta$ and $\cos i\cos\theta$ and not on their individual values. 
The $i-\theta$ degeneracy can be seen in Figure \ref{fig:degeneracy}, where the normalized low-energy band pulse profiles for model A, and another model with $i$ and $\theta$ swapped (both labelled ``Sphere") are indistinguishable. 
For many of the Poisson realizations, the best-fit values for $i$ and $\theta$ shown in Table \ref{tab:Adata} are swapped from their true values, and as a result, the average and standard deviations for the individual angles are really not meaningful, except to illustrate that it is the trigonometric combinations of the angles that can be reliably determined. 
However, since there are independent methods for constraining $i$ and $\theta$ through optical \citep{Wang} or gamma-ray \citep{Venter} observations, it is still useful to discuss these two angles separately.

To ensure that a suitable number of independent realizations of the data were made, a histogram of the best-fit radii is plotted in Figure \ref{fig:Ahistogram}. 
The resulting histogram is approximately Gaussian (70\% are within 1 standard deviation, 95\% within 2 standard deviations, 100\% within 3 standard deviations), indicating that 20 trials are sufficient to illustrate general trends. 
A scatter plot, also shown in Figure \ref{fig:Ahistogram}, shows the values of mass and radius (red crosses) for the 20 different random realizations of the data. 
The large black cross indicates the average and standard deviation values for the mass and radius fits.

{  As previously stated, parameter degeneracy plays a major role in all aspects of uncertainty in our results, even the standard deviation. 
By testing 20 different realizations of the data for each model (a relatively small number),
we still find that some models converge on the true parameter values while others do not. 
The models with stronger parameter degeneracy have larger standard deviations on the parameters.
This provides an initial assessment of which types of models and pulse profile shapes have stronger degeneracies.
Even with very small error bars on the pulse profile, there would still be relatively large standard deviations over many models due to the inherent degeneracies.}

\subsection{Effect of Changing Spot Co-latitude}

In this section, we examine the effect of changing the hotspot's co-latitude $\theta$ on the accuracy and precision of the pulse profile fits. 
By changing $\theta$ while keeping $M$, $R$, and $i$ constant, we alter the projected velocity $\beta$ and the approximate pulse amplitude \hbox{Amp} (see Table 1). 
Models A (with $\theta=20\degrees$), $\theta_{37}$, and $\theta_{60}$ have increasing values of $\theta$, $\beta$, and \hbox{Amp}. 
Due to the $i$-$\theta$ degeneracy, this is also equivalent to keeping $\theta$ fixed and varying $i$.

\begin{figure}[tb]
\epsscale{1.0}
\includegraphics[width=0.49\columnwidth]{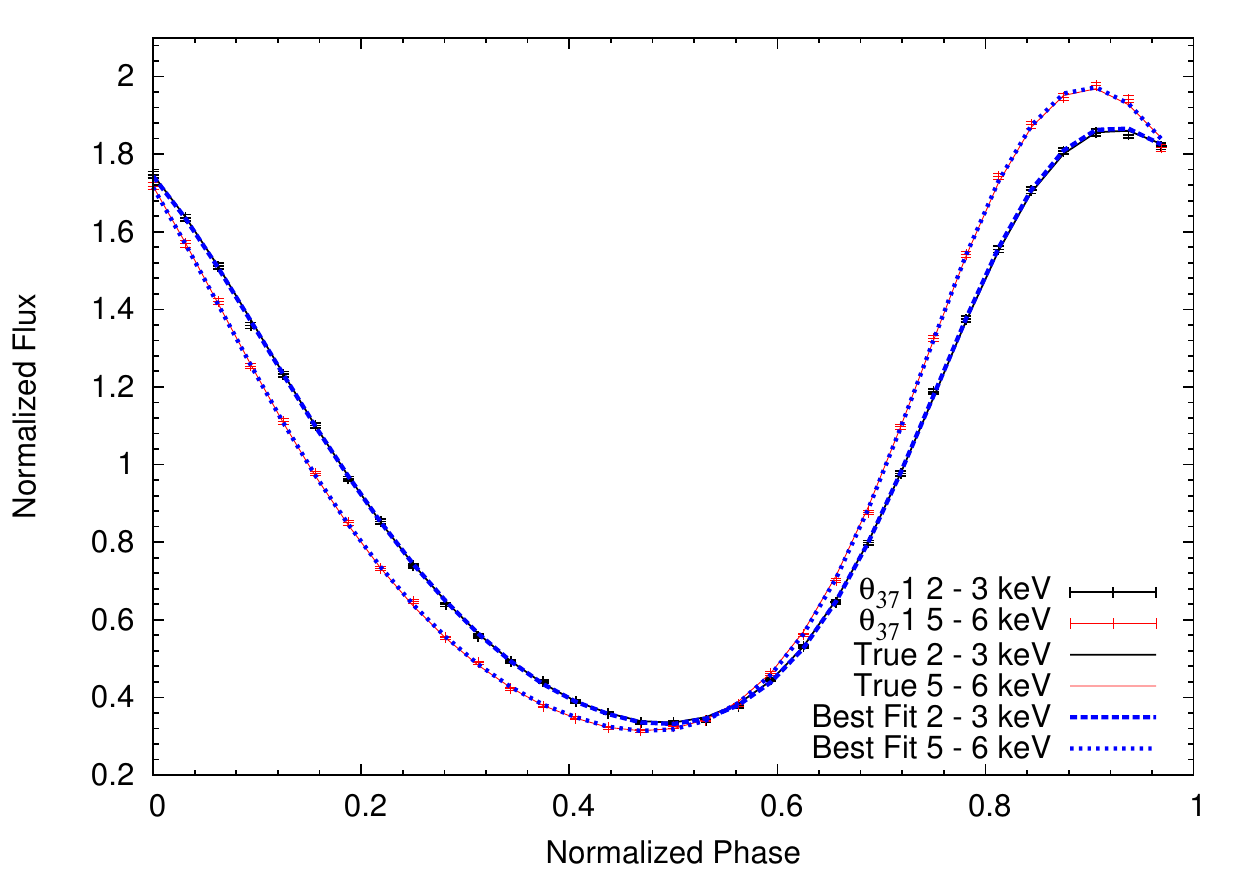}\includegraphics[trim=2cm 6.5cm 2cm 2cm, clip=true, width=0.45\columnwidth]{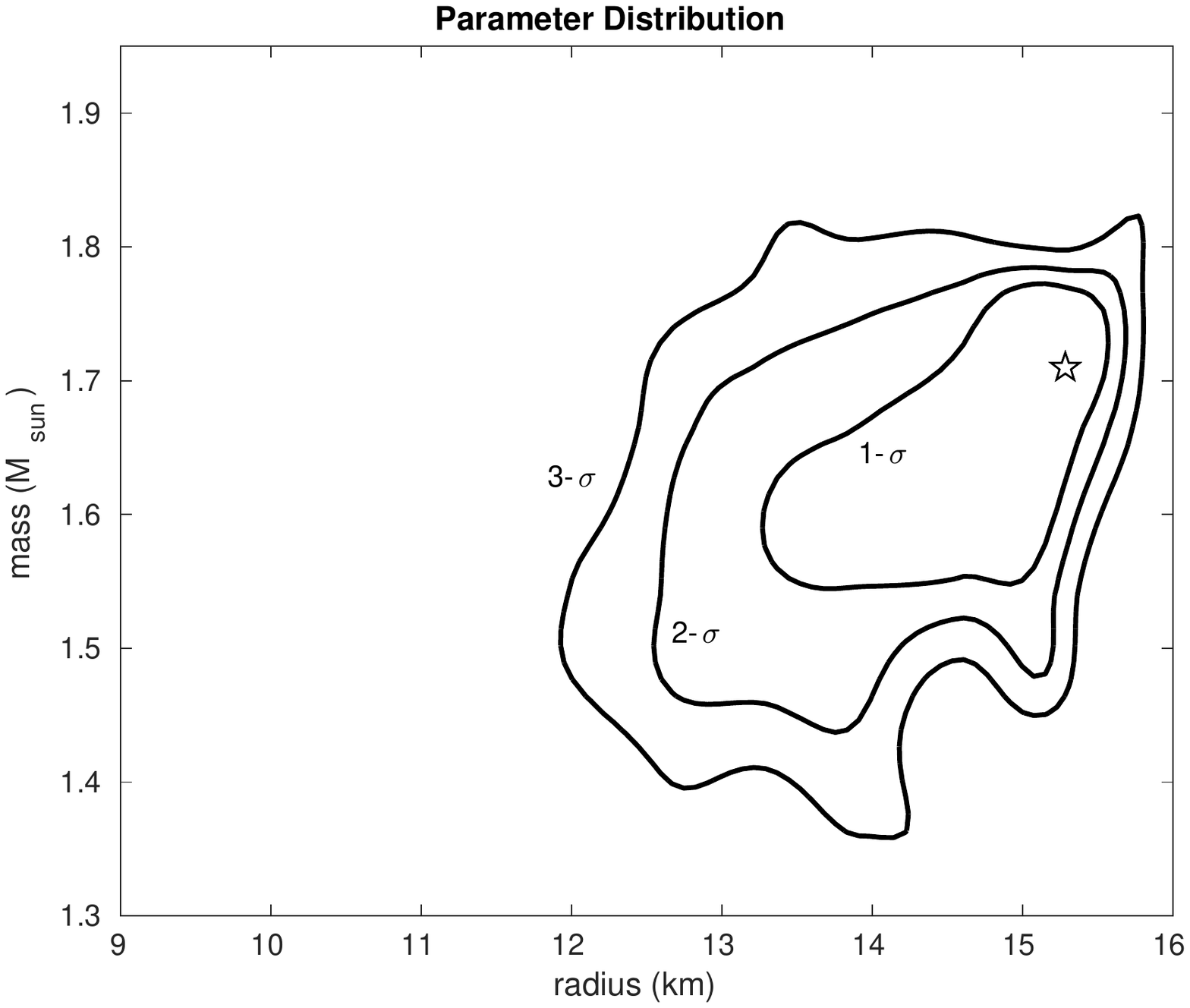}
\caption{\textsc{Left:} Pulse profiles for model $\theta_{37}$ in two energy bands. 
\textsc{Right:} \chis\ contours in the $M-R$ plane for fitting synthetic data from model $\theta_{37}$1.
Symbols and lines have the same meaning as in Figure \ref{fig:A1data}.
The best fit values for mass and radius for this Poisson realization are 1.71 $\Msun$
and 15.29 km, with $\chis=58.7$ for 59 dof.
The approximate 1-\sigdegen\ limits for mass and radius are 0.11 $\Msun$ and 1.1 km.
}
\label{fig:J1data}
\end{figure}

The $\theta_{37}$1 pulse profile (i.e., for  the first Poisson realization of the $\theta_{37}$ model) is shown in the left panel of Figure~\ref{fig:J1data} and the constant \chis\ contours are shown in the right panel. 
This case is one of the ``outlier" pulse shapes (of the 20 Poisson realizations) with a very large best-fit $R$ that only includes the true $R$ at the 3-\sigdegen\ level. 
The 1-\sigdegen\ error regions for the $\theta_{37}$1 model (from the right panel of Figure~\ref{fig:J1data}) are  smaller than the A1 model, close to 1.1 km for $R$ and 0.11 $\Msun$ for $M$. 
The 1-\sigdegen\ regions for this particular Poisson realization are somewhat larger than the standard deviation 1-\sigstat\ computed for the ensemble of $\theta_{37}$ models. 

\begin{figure}[tb]
\epsscale{1.0}
\includegraphics[width=0.49\columnwidth]{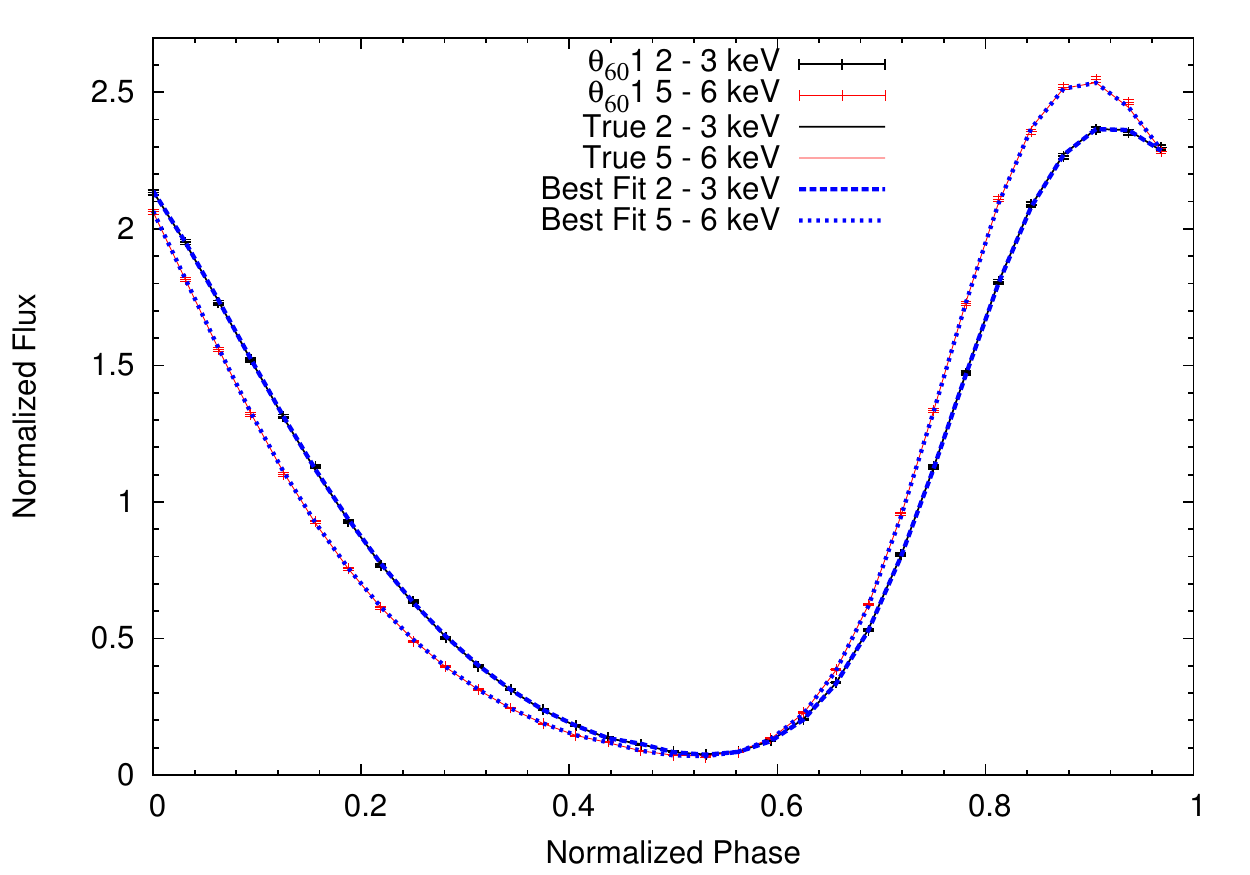}\includegraphics[trim=0.1cm 5cm 0.5cm 0.5cm, clip=true, width=0.40\columnwidth]{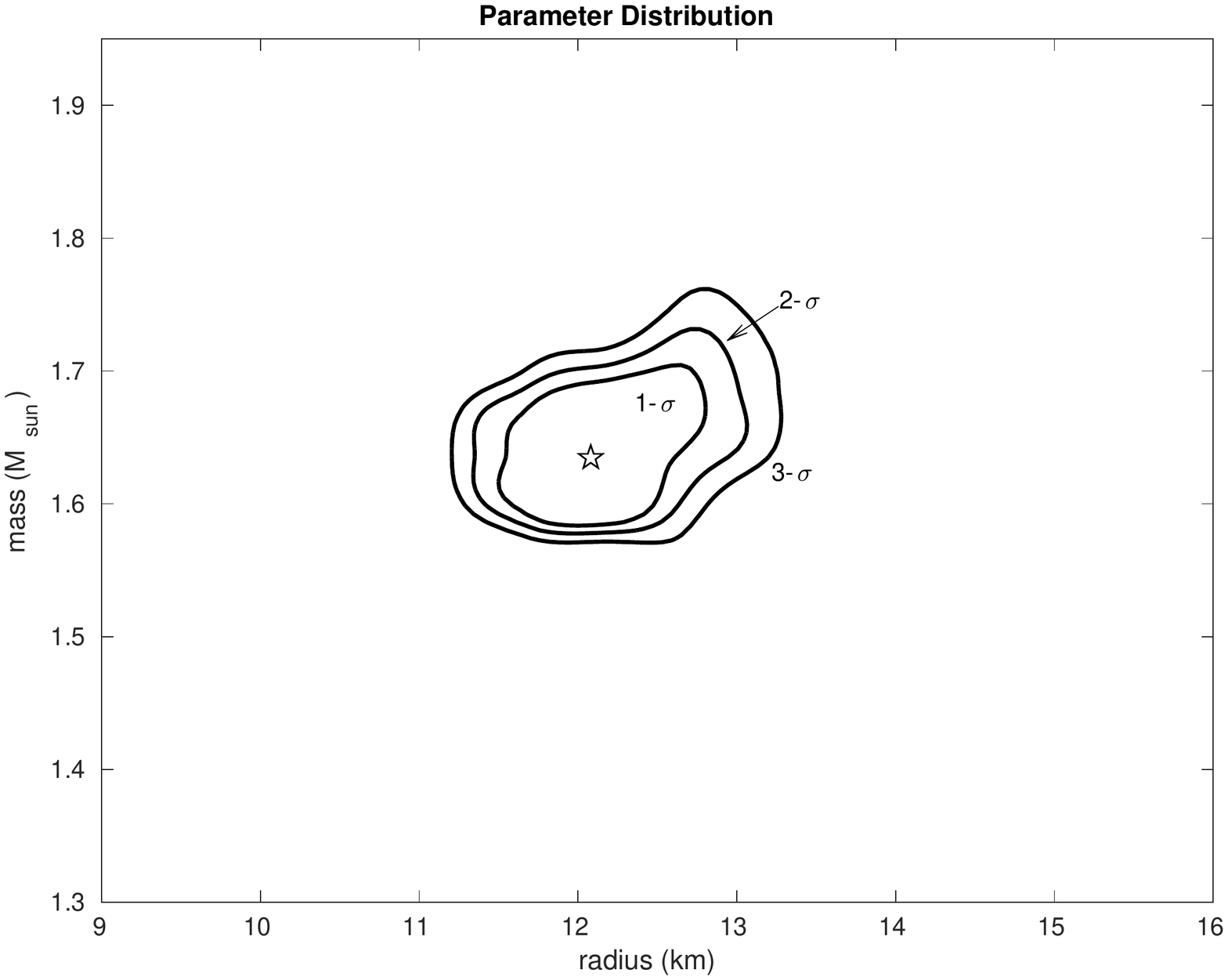}
\caption{\textsc{Left:} Pulse profiles for model $\theta_{60}$ in two energy bands.
\textsc{Right:} \chis\ contours in the $M-R$ plane for fitting synthetic data from model $\theta_{60}$1. 
The symbols have the same meaning as in Figure \ref{fig:A1data}.
The best fit values for mass and radius for this Poisson realization are $M = 1.63 \Msun$
and $R = 12.1$km, with $\chis = 49.6$ for 59 dof. The approximate 1-\sigdegen\ limits for mass and radius are 0.06 $\Msun$ and 0.7 km. 
 }
\label{fig:E1data}
\end{figure}

Similarly, the $\theta_{60}$1 pulse profile (first Poisson realization of the $\theta_{60}$ model) is
shown in the left panel of Figure \ref{fig:E1data} and the constant \chis\ contours are shown in the right panel. 
The 1-\sigdegen\ error regions for the $\theta_{60}$1 model are 0.7 km for $R$ and 0.06 $\Msun$ for $M$, larger than the ensemble standard deviation 1-\sigstat\ by a factor of about 6. 

We find a strong trend as $\theta$ increases: 
the average $M$ and $R$ fit values are closer to the true values and the standard deviation is smaller. 
For example, \sigstat\ in $M$ decreases from 0.11 to 0.08 to 0.01 $M_{\odot}$ for models Models A, $\theta_{37}$, and $\theta_{60}$.
This improvement in fitting accuracy is due to models $\theta_{37}$ and $\theta_{60}$ having a larger \hbox{Amp} and $\beta$ than model A. 
This is consistent with the general trend seen by \citet{Loetal13} in individual model fits.

During an X-ray burst, $\theta$ is expected to change. 
This set of models roughly approximates this process, which has been explored in detail by \citet{Mahmoodifar}.
If a series of \textit{different} burst oscillation pulse shapes are found for a neutron star during the \textit{same} burst, future work could include a simultaneous multi-epoch fit, allowing $\theta$ (and spot size $\rho$)  to be dependent on epoch.

\subsection{Effect of Oblateness} \label{s:oblate}

Rotation alters the shape of a neutron star, making it an oblate spheroid. 
Although the change in shape is small for stars spinning at rates seen in accreting systems, the alteration in the star's shape changes the positions on the neutron star's surface for which photons can reach the observer \citep{Morsinketal07}, leading to large changes in the pulse profile. 
As an example, a 1.6 $M_\odot$ neutron star with an interior given by the APR equation of state \citep{APR} has an equatorial radius of 11.7 km and a ratio of polar to equatorial radii of 0.93. 
This oblate geometry has a larger effect on the pulse profile than other effects due to rotation, such as frame-dragging, as has been discussed in detail in, e.g., \citet{Morsinketal07}. 

In this section, we investigate the effect of the oblate shape on the accuracy and precision of pulse profile fitting and reproducing the input parameters. 
To do this we construct oblate versions of two models, A and $\theta_{37}$. 
The oblate versions, designated with the letter ``O" are constructed so that the radius of the star at the location of the spot is the same as for the corresponding spherical model. 
This means that the values of the star's compactness $M/R$ and the projected velocity $\beta$ are the same for the spherical and oblate models.
As a result of this definition for the radius, the star's equatorial radius is larger than that listed in Table~\ref{tab:Theoretical}. 
The general trend is for the oblate neutron star's pulse profile to have a smaller pulse amplitude than the spherical neutron star with the same parameters (when $i$ and $\theta$ are in the same hemisphere), since the visibility condition makes it easier to see the far side of the neutron star when the star is oblate \citep{Cadeauetal07, Morsinketal07}. 
In cases where the spot is visible for all phases (as in both models A and $\theta_{37}$), consider the emission when the spot and the observer have the same azimuthal angle. 
At this moment, the light from the spherical star is emitted close to the normal to the surface. 
For the oblate star, the light is emitted in the same direction in space, but it is at an angle further from the surface's normal, due to the tilt of the surface. 
Since the intensity of light is proportional to the cosine of the angle between the original direction of emission and the normal to the surface, the spherical star's spot appears brighter at this phase. 
The opposite is true when the spot is on the opposite side of the star from the observer.
In this case the light is emitted close to the tangent to the surface for the spherical star, and closer to the normal for the oblate star, making the spot appear dimmer for the spherical star at this phase. 
The overall effect is to create a less modulated pulse profile for the oblate star. 

In model AO, the neutron star's equatorial radius is 12.7\,km, while the radius at the spot's latitude is 12\,km. 
As a comparison with the spherical model, the low-energy band pulse profiles for both AO and A are plotted as black dot-dashed and solid curves respectively in Figure~\ref{fig:degeneracy}. 
The high energy band (not shown for clarity) has a similar decrease in modulation.  
In model $\theta_{37}$O the equatorial radius is 12.5\,km, and has a lower pulse amplitude than model $\theta_{37}$. 
Since the pulse profiles and $M-R$ confidence regions for models AO and $\theta_{37}$O look quite similar to those for models A and $\theta_{37}$, they are not shown here.
The mean and standard deviation of the best fit parameters  for the 20 realizations for each oblate model are given in Table \ref{tab:summary}. 
For most parameters, the \sigstat\ and accuracies are smaller for the oblate case than for the fiducial spherical case (we also found that the 1-\sigdegen\ regions were smaller by about a half).
A similar improvement in the precision of the results was also noted by \citet{Miller15}, however since they only considered one Poisson realization of the data they could not rule out the possibility that this was due to a statistical fluctuation. 
In our case, since we are comparing a sample of Poisson realizations for each model, the increase in precision and accuracy is most likely due to the properties of the oblate model pulse profiles. 

The improvement in the accuracy and precision in the determination of most of the parameters for oblate models is most likely due to the partial lifting of the degeneracy between $i$ and $\theta$. 
For an oblate star, the direction that the normal to the surface points in depends on the shape of the star which is a function of $\theta$, but is independent of $i$. 
This introduces small differences between the normalized pulse profiles for models with $i$ and $\theta$ switched, while for spherical stars, the normalized pulse profiles are the same when the angles are switched. 
This is illustrated in Figure~\ref{fig:degeneracy} where the two oblate models representing the swapped inclination and spot angles, shown with black dot-dashed and red dotted curves, are clearly different from each other. 
However, since they are still fairly similar to each other, there is still a partial degeneracy when the angles are swapped. 

{  There is a similar but much smaller lifting of degeneracy for large hot spots on spherical stars. A large spot extends over a range of $\theta$ values while the observer's 
inclination $i$ is just one fixed value, so swapping the two angles if the spot is large will not yield the same pulse shape. However, the magnitude of the effect is much smaller than the magnitude of the change that occurs when the angles are swapped on an oblate star. }

We also tested the effect of using the wrong shape model by using the oblate data corresponding the 20 AO models as input and fitting them with pulse shapes for spherical stars.
In this case, \chis\ increased a small amount but the best-fit values for $M$ and $R$ became very inaccurate. 
For the set of fits to the Poisson realizations, the average fit returned $\chis=64.7$, $M=1.91\Msun$, and $R=12.3\,$km (which should be compared with the values in the 4th row of Table~\ref{tab:summary}). 
The average $M$ and $R$ fits are accurate to 19.4\% and 2.5\%, respectively.
Similar results were found by \citet{Miller15}.
In practice, this is not a problem, since we know that the stars are actually oblate, so the correct shape can be used. 
{  The spherical shape model is computationally somewhat cheaper, hence it is used for the other models tested in this paper.}
Furthermore, the oblate shape model used is still an approximation.
It would be worthwhile, in future research, to compare the slightly different shape models that have been used in this paper with other models used by other groups 
(such as \citealt{Baubock13, Miller15}).

\subsection{Effect of Photon Count-Rate} \label{s:noise}

The average photon count-rate per phase bin of 25000 is the expected best case scenario, assuming typical burst flux, large detector effective area, and long observation time \citep{Ferocietal12},
and similar to the count-rate used in other pulse profile analyses \citep{Loetal13,Psaltis14b}.
The pulse profile is noisier when fewer counts are detected.
We tested the effect of statistical noise on model $\theta_{60}$ by generating different Poisson realizations, assuming a reduced count-rate, of the same theoretical pulse profile.
Comparing model $\theta_{60}$ with the reduced count-rate model, $\theta_{60}C_{6250}$, allows us to explicitly compare the {  degeneracy-related uncertainty} versus error.
The average counts per bin are 25000 and 6250 for models $\theta_{60}$ and $\theta_{60}C_{6250}$, respectively. 

The pulse profile for model $\theta_{60}C_{6250}$ looks very much like that for model $\theta_{60}$ but with larger errors, so it is not shown.
The \chis\ contours for  model $\theta_{60}C_{6250}1$ are shown in Figure \ref{fig:Ecdata}. 
The size of the 1-\sigdegen\ contours in Figures \ref{fig:E1data} and \ref{fig:Ecdata} are very similar, suggesting that the confidence contours for one particular mock observation are mainly dominated by the parameter degeneracy.

From Table~\ref{tab:summary}, the standard deviations of the parameters for the low count-rate model are $\simeq2-3$ times as large as those of model $\theta_{60}$. 
The factor of two is expected purely on statistics (from the factor 4 reduction in counts).
The actual degradation is somewhat worse, likely because partial degeneracy between the parameters makes the parameter determination degrade more rapidly than the errors. Model $\theta_{60}C_{6250}$ still gives well-determined parameters even at the lower count-rate value. 
The accuracies in $M$ and $R$ determination are 2\%  and 4\% respectively, for the low count-rate case. 

\begin{figure}[tb]
\centering
\epsscale{1.0}
\includegraphics[trim=0.1cm 6cm 0.5cm 0.5cm, clip=true, width=0.45\columnwidth]{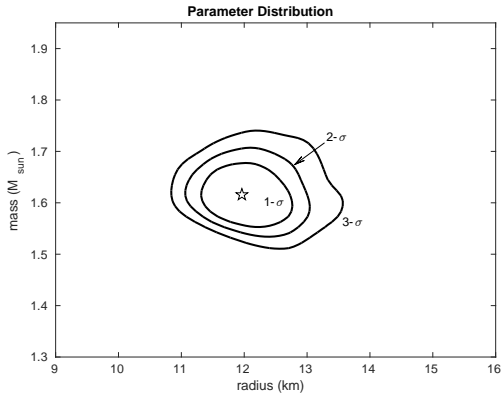}
\caption{ \chis\ contours in the $M-R$ plane for fitting synthetic data from model $\theta_{60}C_{6250}$1. 
The symbols have the same meaning as in Figure \ref{fig:A1data}.
The best fit values for mass and radius for this Poisson realization are $M=1.62\Msun$ and $R=12.0$\,km, with $\chis=53.8$ for 59 dof.
The approximate 1-\sigdegen\  limits for mass and radius are 0.06 $\Msun$ and $0.8$ km. 
}
\label{fig:Ecdata}
\end{figure}

\subsection{Effect of Compactness} \label{s:beta}

Changes in the compactness ratio ($M/R$) are expected to impact the quality of the fits, since the compactness affects the pulse amplitude \hbox{Amp} (evident in equation (\ref{eqn:amp})). 
Increasing the compactness decreases \hbox{Amp}, which might be expected to decrease the precision and/or accuracy of the fits. 
To test this effect, we generated models with differing values for the compactness ratio, but the same values of the projected spot velocity $\beta$. 

We created two models similar to model A with a larger and smaller compactness ratio, $(M/R)_{hi}$ and $(M/R)_{lo}$, and generated a set of pulse shapes with Poisson noise.
The results of the fits can be seen in Table~\ref{tab:summary}. 
All three models with the same value of $\beta$ have similar accuracy and precision for most of the parameters. 
For instance, the accuracy of determining the mass ranges from $4-5\%$, while the precision ranges from $6-7\%$, so the change in compactness does not greatly affect the fits. 
However, while there is little change in the precision of the radius measurement ($9-10\%$), in the case of the low compactness model the accuracy was improved.

As a further test on the effect of compactness, model $\beta_{hi}(M/R)_{hi}$ was created to compare with the high spot co-latitude model $\theta_{60}$.
The parameters were the same except that $\theta$ was adjusted to give the same $\beta$ as model $\theta_{60}$.
This also necessitated that \hbox{Amp} was somewhat lower than model $\theta_{60}$ (see equations (\ref{eqn:amp})  and (\ref{eqn:beta})). 
Previously for model $\theta_{60}$, the accuracy and precision for both $M$ and $R$ were better than 1\%; 
by fitting model $\beta_{hi}(M/R)_{hi}$, we found that increasing the compactness degrades both the accuracy and precision in $M$ and $R$, but only so that they range from 1-3\%. 

From these results, given a value of $\beta$, it appears that the accuracy and precision to which the other parameters can be determined does not strongly depend on the compactness ratio of the star. 
Thus, our fit results would not be significantly different if we had chosen a different $M$ and $R$ for the fiducial model A.

\subsection{Effect of Spin Frequency} \label{s:frequency}

The last parameter we adjusted to be different from the fiducial model was $\nu_\text{spin}$. 
We compared model A ($\nu_\text{spin}=600\,$Hz) with a 400 Hz model, labelled $\nu_{400}$. 
Model $\nu_{400}$ has all parameters the same as model A except $\nu_\text{spin}$,
and hence a different projected spot velocity ($\beta$) (see Table \ref{tab:Theoretical}). 
This model has parameter values very similar to the ``low inclination" model presented by \citet{Loetal13}.

The fits results are summarized in Table \ref{tab:summary}.
The precision (standard deviations) and accuracy (difference of mean and true values) are worse for model $\nu_{400}$, as expected. 
For $M$, $R$, $M/R$, and Amp, the precisions are about twice as large for the 400 Hz model
as for the standard 600 Hz model.
However, $\beta$ has the same absolute precision for the 400 Hz model as for the 600 Hz model.

\subsection{Effect of Incorrect Model or Additional Free Parameters} \label{s:incorrect}

Along with the sets of trials summarized in the previous subsections, we tested some of the model assumptions used in the fits. 
These tests include the assumptions we have made about the spectral emissivity, spot size, presence of a constant background flux, and the star's shape. 
Errors are not reported for the tests of this subsection since fits were only carried out on the A1 synthetic data set, the first Poisson realization of the Model A theoretical curve.

We first tested the effect of using the wrong atmosphere model in the fits. 
To do this, we used the synthetic data corresponding to the A1 model (see Figure~\ref{fig:A1data}) as input to Ferret. 
The synthetic data were generated using the Hopf limb-darkening function, but Ferret was run assuming a perfectly isotropic blackbody spectrum. 
The algorithm was unable to find a good fit after 300 generations (about double the number of generations normally required) and the best fit had $\chis= 316.6$ for 59 dof. 
The best-fit $M$ and $R$ were $1.21\Msun$ and $11.3\,$km, which are quite inaccurate (24.4\% for $M$, 5.8\% for $R$). 
As was seen by \citet{Loetal13}, one does not get a good fit using an incorrect atmosphere model.

We next tested fitting a variable spot size model to a pulse profile that was generated with an infinitesimally small spot. 
The infinitesimal size of the spot is a necessary assumption due to the large number of trials carried out in this work. 
In order to model a larger spot, the spot has to be cut into a number of segments, and separate computations of the deflection angles must be computed, increasing the run time linearly with the number of segments.
The pulse shape for a large spot tends to have a lower pulse amplitude than a small spot centred at the same latitude, due to the effect of averaging over many latitudes. 
As a test of this assumption, we used the A1 data set as input to the Ferret algorithm, but added the angular radius $\rho$ of the spot as a free parameter that was allowed to vary between 1\degrees and 60\degrees. 
The resulting best fit has $\chis = 59$ (for 58 dof) and converged on an angular radius of $\rho=1$\degrees, the smallest angle allowed. 
The best-fit $M$ and $R$ for this case are $1.51\Msun$ and $11.6$ km. 
While these values are less accurate (5.6\% and 3.3\%, respectively) than the best-fit value shown in the first row of Table~\ref{tab:Adata} corresponding to a fit with an infinitesimal spot, they are within the 1-\sigstat\ limits computed for the set of model A fits. 
Since using multiple spot segments leads to a significantly longer computation time, it was not feasible to use variable spot sizes in this paper. 
However, in future work when real data are being fitted, it will be necessary to allow the spot size to vary, and to compute the observed flux from multiple spot segments.

We also tested the effect of allowing the temperature of the spot to be a free parameter in the fitting program. 
The theoretical model was constructed with a spot temperature of 2 keV (as measured {  in the comoving frame} at the star's surface). 
The Ferret program was used to fit the data with the addition of a local temperature parameter that was allowed to vary between 1 and 3 keV. 
The best fit returned $\chis= 57.9$ (for 58 dof), $M=1.67\Msun$, $R=11.8$\,km, and a temperature of 2.1 keV. 
The accuracies in $M$ and $R$ are 4.8\% and 1.7\%, respectively.
Although these $M$ and $R$ values are within the 1-\sigstat\ limits for this data set, the addition of the extra parameter does degrade the accuracy. 
This is most likely due to the degeneracy introduced since we measure the redshifted temperature with the light curves.
For the case of real data, it would be important to allow the temperature to vary in the fits, {  since the temperature of the spot in the comoving frame is unknown.
Furthermore, making} use of multiple energy bands would improve the accuracy {  of the temperature measured at infinity. }

The effect of an unknown background flux has been investigated in detail by \citet{Loetal13}, and for comparison we test it for one simple case. 
The synthetic data set A1 was constructed {  without a background count rate.} 
We used this as input and allowed Ferret to add two new parameters corresponding to a constant (in time) background flux in each energy band. 
These background values were allowed to vary between 0 and 1 (recall that our pulse profiles are normalized to 1). 
The resulting fit had $\chis= 57.7$ (for 57 dof), and the best-fit $M$ and $R$ were 1.52$\Msun$ and 11.7\,km. 
$M$ and $R$ were accurate to 5.0\% and 2.5\%, respectively. 
It is possible to add a more realistic background model by adding emission at a lower temperature from the rest of the star \citep{Loetal13,Elshamouty} or by adding light scattered from the disk \citep{MorsinkLeahy11}, but at this point it is not obvious what the most realistic model would be. 
{  Real data are also likely to contain} non-negligible emission from the surface of the neutron star outside the spot region and from the accretion disk.

The effect of these tests on the best-fit values of $M$ and $R$ are summarized in Table~\ref{tab:A1tests}. 
These tests highlight the importance of having as realistic an atmosphere model as possible, since fitting with the wrong model drastically affected the quality and accuracy of the fits. Thus, this method provides
a sensitive test for atmosphere models.
Adding a free parameter for the temperature, unknown background flux, or spot size did not significantly detract from the accuracy in $M$ and $R$, and our code was able to replicate the additional parameter quite well.
Ideally, for real data we would allow these parameters to also vary in the fits, even if it is not feasible to do so in the present study.

\section{Discussion and Conclusion}
\label{s:discussion}

We calculated pulse profiles for simulated thermonuclear burst oscillations from a rapidly rotating neutron star as would be detected by a next-generation X-ray timing observatory, such as ASTROSAT, NICER, or LOFT. 
The input neutron star parameters include mass $M$, radius $R$, emitting spot co-latitude $\theta$, and observer inclination $i$. 
We created 20 Poisson realizations of each test pulse profile, which allows us to determine the errors in the derived model parameters, focusing on $M$ and $R$.
The resulting pulse profiles were fitted with Ferret to analyze how well the input parameters could be recovered (in both standard deviation and {  degeneracy-related uncertainty}). 

We find that the best-determined parameter is the projected velocity $\beta$ of the spot along the observer's line-of-sight.
The next best-determined are $M$ and $R$. 
Compactness (M/R), $\sin i \sin\theta$, and the pulse amplitude \hbox{Amp} are also well-determined, but $\cos i \cos\theta$, $i$, and  $\theta$ are poorly determined. 
It is clear that more rapidly rotating neutron stars produce pulse profiles with strong harmonics, which yield better constraints on their parameters. 
Also, neutron stars viewed at larger $i$ with spots at a larger $\theta$ produce more modulated pulse profiles and therefore better constraints. 
However, we also found that count-rate is important, so the best targets will be those systems which have the optimal combination of large $\nu_\text{spin}$, large $\theta$ and $i$, and bright pulsations. 
For our best cases presented here, $M$ and $R$ were determined to 1\% accuracy and precision, but even for many of the less optimum cases $M$ and $R$ were determined to $\sim$5\% accuracy and precision, which is a very valuable result.
 
We carried out a number of parameter comparison tests, to see how different input parameters can affect the constraints on $M$ and $R$. 
The more asymmetry and larger pulse amplitude, the better the accuracy and precision in $M$ and $R$ from pulse profile fitting. 
The asymmetry in the pulse profile is controlled by $\beta$; a larger $\beta$ results from increases in $\theta$, $i$, and $\nu_\text{spin}$. 
A larger $\theta$ or $i$ also gives a larger \hbox{Amp}.
Compactness ($M/R$) has a small effect on parameter determination; 
for more compact stars it is somewhat more difficult to determine parameters. 
This is caused by the increased visibility of the surface by the observer and the resulting decreased pulse amplitude.
Including oblateness in the pulse profile model improves the accuracy and precision of $M$ and $R$ determinations. 
This is due to a reduced $i-\theta$ degeneracy in the oblate models. 

Photon count-rate has a critical effect on parameter determination. 
On simple grounds, increasing the counts by a factor of four reduces the error by a factor of two. 
However, in practise we found a factor of three improvement in parameter determination. 
The extra gain is likely the result of reduced parameter degeneracy for data with smaller errors. 

We assumed the spot temperature to be a known definite quantity. 
In practice, this is determined by a spectral fit to the data using multiple energy bands. 
We calculated the simulated pulse profiles using only two energy bands, so we have underestimated the uncertainties in $M$ and $R$ that enter through the uncertainty of the temperature. 
In principle, it is not difficult to use more energy bands (e.g., \citealt{Loetal13}). However, there is a trade off that the signal-to-noise in each energy band falls as more, and thus smaller, energy bands are used. 
Determining an optimal number of energy bands that allows a determination of the spectrum while providing enough statistics to constrain the star's properties should be the topic of a future study.

We also took the background count rate to be negligible in comparison with the count rate from the neutron star hot spot. 
For a real observation, the background comes from three sources: instrument background, sky background, and source contribution to background. 
Instrument background depends strongly on the instrument design and mode of observation.
Source background can consist of emission from the surface of the neutron star outside of the spot region and from the accretion disk. 
{  However, subtracting the persistent pre-burst emission from the X-ray burst flux is not an appropriate source-background subtraction. 
As shown in \citet{Worpeletal13}, the persistent emission can be different during the burst, possibly due to an increased accretion rate onto the neutron star. 
A method such as weighted-photons Bayesian Blocks \citep{Worpeletal15} should be incorporated into the analysis pipeline to subtract the appropriate level of source-background emission.

A limitation of our study is that we did not include the effect of frequency drift, which is normally seen during the rise of an X-ray burst \citep{Watts12}. 
When dealing with real data, it would be important to either remove sections of the data where frequency drift is observed, or to use a reliable method for modelling the pulse shape with phase offsets to account for the drift. 
An additional complication for the sources that also have accretion-powered pulsations is that the pulsations may contaminate the X-ray burst oscillations. 
At this time these are open problems associated with modelling X-ray burst oscillations. 

In this paper we have demonstrated that an evolutionary optimization and parameter search methods
 can be used to effectively constrain the mass and radius of a neutron star with pulsed emission. 
We focussed on burst oscillations seen in Type I X-ray bursts, however the routines can also be used to model accretion-powered pulsations or rotation-powered pulsations once the appropriate spectrum and beaming functions have been changed. 
In particular, the rotation-powered pulsars that will be studied by NICER (see \citealt{Ozel15}) can be analyzed using the methods discussed in this paper.

\acknowledgments
This research was supported by grants from the Natural Sciences and Engineering Research Council of Canada to J.D.F., D.A.L., and S.M.M. 
A.L.S. acknowledges a travel grant from the LKBF (Leids Kerkhoven-Bosscha Fonds). 
We thank Arash Bahramian, Frederick Lamb, M. Coleman Miller, Adam Rogers, and Christoph Weniger for useful discussions. 
We also thank the LOFT Dense Matter science working group for early code comparisons to validate the modelling.

\bibliographystyle{apj}

\begin{table*} 
\centering
\caption{Summary of Best Fits for Model A \label{tab:Adata}}
\begin{tabular}{rccccccccccc}
\toprule
Model & \chis & $M$ & $R$ & $i$ & $\theta$ & $\phi$ & $M/R$ & $\sin i \sin\theta$ & $\cos i \cos\theta$ & Amp & $\beta$\\
& 59 dof & \Msun & km & deg. & deg. & $\times 10^{-3}$ & & & & & $\times10^{-2}$ \\
\midrule[\heavyrulewidth]
True & &  1.600&  12.00&   60.0& 20.0&  0.0000& 0.1969& 0.296&  0.470&  0.373& 5.741 \\ 
\midrule  
1&    58.2& 1.575&  11.97&   48.7& 25.0&  0.0071& 0.1943& 0.318&  0.598&  0.347& 6.120 \\ 
2&    60.8& 1.517&  12.39&   19.9& 57.3&  -0.0011& 0.1808& 0.286&  0.508&  0.362& 5.577 \\ 
3&    58.6& 1.344&  9.39&   39.4& 34.4&  0.0021& 0.2114& 0.358&  0.638&  0.357& 5.565 \\ 
4&    71.7& 1.532&  10.70&   21.7& 59.8&  0.0000& 0.2114& 0.319&  0.467&  0.383& 5.658 \\ 
5&    63.9& 1.666&  11.46&   71.8& 17.6&  -0.0062& 0.2147& 0.287&  0.297&  0.426& 5.471 \\ 
6&    56.7& 1.445&  10.26&   36.7& 36.0&  0.0064& 0.2080& 0.351&  0.649&  0.350& 5.932 \\ 
7&    64.6& 1.469&  10.81&   37.2& 34.4&  0.0074& 0.2007& 0.342&  0.657&  0.345& 6.007 \\ 
8&    53.9& 1.694&  12.39&   68.7& 17.3&  -0.0047& 0.2019& 0.277&  0.347&  0.404& 5.584 \\ 
9&    42.3& 1.417&  11.76&   42.8& 27.5&  0.0044& 0.1779& 0.313&  0.651&  0.338& 5.773 \\ 
10&    67.0& 1.510&  10.32&   21.9& 60.6&  -0.0007& 0.2161& 0.324&  0.456&  0.388& 5.583 \\ 
11&    57.1& 1.581&  11.29&   21.7& 58.1&  0.0026& 0.2068& 0.314&  0.491&  0.372& 5.828 \\ 
12&    68.9& 1.433&  12.08&   29.1& 39.9&  0.0062& 0.1752& 0.311&  0.671&  0.331& 5.869 \\ 
13&    58.4& 1.458&  10.88&   35.6& 35.8&  0.0075& 0.1979& 0.341&  0.659&  0.345& 5.994 \\ 
14&    63.7& 1.702&  13.86&   15.9& 67.6&  -0.0057& 0.1813& 0.254&  0.366&  0.390& 5.541 \\ 
15&    58.7& 1.447&  9.59&   38.1& 36.9&  0.0067& 0.2228& 0.371&  0.629&  0.360& 6.006 \\ 
16&    45.5& 1.746&  13.31&   18.3& 63.1&  0.0015& 0.1937& 0.280&  0.429&  0.375& 5.983 \\ 
17&    51.8& 1.435&  12.96&   30.7& 36.3&  0.0086& 0.1635& 0.302&  0.693&  0.323& 6.009 \\ 
18&    53.1& 1.488&  10.57&   38.4& 34.2&  0.0076& 0.2079& 0.349&  0.648&  0.348& 6.076 \\ 
19&    72.0& 1.591&  12.22&   19.9& 59.0&  0.0003& 0.1923& 0.292&  0.485&  0.367& 5.727 \\ 
20&    58.1& 1.438&  9.68&   38.5& 36.1&  0.0060& 0.2195& 0.367&  0.632&  0.359& 5.962 \\ 
\midrule
Ave&  59.3 & 1.524&  11.39&   34.7& 41.8& 0.0028  & 0.1989& 0.318&  0.549&  0.363& 5.813 \\ 
Std& 7.6 &   0.107&  1.23&   15.0& 15.2& 0.0046  & 0.0159& 0.032&  0.120&  0.025& 0.203 \\
\bottomrule
\end{tabular}
\\ The summary of fits for the other models are available online only.
\vspace{12pt}
\end{table*}

\begin{table*} 
\centering
\footnotesize
\caption{Summary of Tests on A1 Data \label{tab:A1tests}}
\begin{tabular}{lrcccccccccc}
\toprule
Model & $\chis/\text{dof}$ & $M$ & $R$ & $i$ & $\theta$ & $\phi$ & $M/R$ & $\sin i \sin\theta$ & $\cos i \cos\theta$ & Amp & $\beta$\\
 & & \Msun & km & deg. & deg. & $\times 10^{-3}$ & & & & & $\times10^{-2}$ \\
\midrule[\heavyrulewidth]
True& &  1.600&  12.00&   60.0& 20.0&  0.0000& 0.1969& 0.296&  0.470&  0.373& 5.741 \\ 
\midrule   
A1 Best Fit &    $58.2/59$ & 1.575&  11.97&   48.7& 25.0&  0.0071& 0.1943& 0.318&  0.598&  0.347& 6.120 \\ 
Wrong Atmosphere&    $316.6/59$ & 1.210&  11.29&   40.2& 39.2&  0.0104& 0.1583& 0.408&  0.592&  0.495& 7.004 \\ 
Spot Size Varies&    $59.0/58$ & 1.508&  11.57&   33.3& 37.0&  0.0090& 0.1925& 0.331&  0.667&  0.338& 6.137 \\ 
Temperature Varies&    $57.9/58$ & 1.668&  11.81&   23.1& 55.1&  0.0057& 0.2086& 0.322&  0.526&  0.364& 6.266 \\ 
Background &    $57.7/57$ & 1.522&  11.75&   45.8& 26.7&  0.0066& 0.1913& 0.321&  0.623&  0.344& 6.045 \\
\bottomrule
\end{tabular}
\vspace{12pt}
\end{table*}

\end{document}


\begin{table*} 
\centering
\caption{Summary of Best Fits for Model A \label{tab:Adata}}
\begin{tabular}{rccccccccccc}
\toprule
Model & \chis & $M$ & $R$ & $i$ & $\theta$ & $\phi$ & $M/R$ & $\sin i \sin\theta$ & $\cos i \cos\theta$ & Amp & $\beta$\\
& 59 dof & \Msun & km & deg. & deg. & $\times 10^{-3}$ & & & & & $\times10^{-2}$ \\
\midrule[\heavyrulewidth]
True & &  1.600&  12.00&   60.0& 20.0&  0.0000& 0.1969& 0.296&  0.470&  0.373& 5.741 \\ 
\midrule  
1&    58.2& 1.575&  11.97&   48.7& 25.0&  0.0071& 0.1943& 0.318&  0.598&  0.347& 6.120 \\ 
2&    60.8& 1.517&  12.39&   19.9& 57.3&  -0.0011& 0.1808& 0.286&  0.508&  0.362& 5.577 \\ 
3&    58.6& 1.344&  9.39&   39.4& 34.4&  0.0021& 0.2114& 0.358&  0.638&  0.357& 5.565 \\ 
4&    71.7& 1.532&  10.70&   21.7& 59.8&  0.0000& 0.2114& 0.319&  0.467&  0.383& 5.658 \\ 
5&    63.9& 1.666&  11.46&   71.8& 17.6&  -0.0062& 0.2147& 0.287&  0.297&  0.426& 5.471 \\ 
6&    56.7& 1.445&  10.26&   36.7& 36.0&  0.0064& 0.2080& 0.351&  0.649&  0.350& 5.932 \\ 
7&    64.6& 1.469&  10.81&   37.2& 34.4&  0.0074& 0.2007& 0.342&  0.657&  0.345& 6.007 \\ 
8&    53.9& 1.694&  12.39&   68.7& 17.3&  -0.0047& 0.2019& 0.277&  0.347&  0.404& 5.584 \\ 
9&    42.3& 1.417&  11.76&   42.8& 27.5&  0.0044& 0.1779& 0.313&  0.651&  0.338& 5.773 \\ 
10&    67.0& 1.510&  10.32&   21.9& 60.6&  -0.0007& 0.2161& 0.324&  0.456&  0.388& 5.583 \\ 
11&    57.1& 1.581&  11.29&   21.7& 58.1&  0.0026& 0.2068& 0.314&  0.491&  0.372& 5.828 \\ 
12&    68.9& 1.433&  12.08&   29.1& 39.9&  0.0062& 0.1752& 0.311&  0.671&  0.331& 5.869 \\ 
13&    58.4& 1.458&  10.88&   35.6& 35.8&  0.0075& 0.1979& 0.341&  0.659&  0.345& 5.994 \\ 
14&    63.7& 1.702&  13.86&   15.9& 67.6&  -0.0057& 0.1813& 0.254&  0.366&  0.390& 5.541 \\ 
15&    58.7& 1.447&  9.59&   38.1& 36.9&  0.0067& 0.2228& 0.371&  0.629&  0.360& 6.006 \\ 
16&    45.5& 1.746&  13.31&   18.3& 63.1&  0.0015& 0.1937& 0.280&  0.429&  0.375& 5.983 \\ 
17&    51.8& 1.435&  12.96&   30.7& 36.3&  0.0086& 0.1635& 0.302&  0.693&  0.323& 6.009 \\ 
18&    53.1& 1.488&  10.57&   38.4& 34.2&  0.0076& 0.2079& 0.349&  0.648&  0.348& 6.076 \\ 
19&    72.0& 1.591&  12.22&   19.9& 59.0&  0.0003& 0.1923& 0.292&  0.485&  0.367& 5.727 \\ 
20&    58.1& 1.438&  9.68&   38.5& 36.1&  0.0060& 0.2195& 0.367&  0.632&  0.359& 5.962 \\ 
\midrule
Ave&  59.3 & 1.524&  11.39&   34.7& 41.8& 0.0028  & 0.1989& 0.318&  0.549&  0.363& 5.813 \\ 
Std& 7.6 &   0.107&  1.23&   15.0& 15.2& 0.0046  & 0.0159& 0.032&  0.120&  0.025& 0.203 \\
\bottomrule
\end{tabular}
\vspace{12pt}
\end{table*}

\begin{table*} 
\centering
\caption{Summary of Best Fits for Model AO \label{tab:AOdata}}
\begin{tabular}{rccccccccccc}
\toprule
Model & \chis & $M$ & $R$ & $i$ & $\theta$ & $\phi$ & $M/R$ & $\sin i \sin\theta$ & $\cos i \cos\theta$ & Amp & $\beta$\\
& 59 dof & \Msun & km & deg. & deg. & $\times 10^{-3}$ & & & & & $\times10^{-2}$ \\
\midrule[\heavyrulewidth]
True&&       1.600&  12.00&   60.0& 20.0&  0.000& 0.1969& 0.296&  0.470&  0.373& 5.741 \\ 
\midrule 
1&    58.8& 1.643&  12.48&   34.2& 32.9&  0.0114& 0.1944& 0.306&  0.694&  0.302& 6.140 \\ 
2&    58.7& 1.590&  11.79&   63.0& 19.2&  -0.0034& 0.1992& 0.293&  0.429&  0.385& 5.600 \\ 
3&    58.0& 1.721&  11.57&   20.9& 52.1&  -0.0030& 0.2197& 0.282&  0.573&  0.292& 5.479 \\ 
4&    70.8& 1.584&  11.00&   42.9& 28.0&  0.0040& 0.2127& 0.319&  0.647&  0.314& 5.830 \\ 
5&    64.0& 1.756&  11.29&   21.5& 52.2&  -0.0004& 0.2297& 0.290&  0.570&  0.292& 5.604 \\ 
6&    51.9& 1.588&  9.52&   57.4& 23.4&  -0.0030& 0.2463& 0.335&  0.494&  0.342& 5.629 \\ 
7&    65.3& 1.624&  11.36&   60.2& 20.5&  -0.0010& 0.2111& 0.304&  0.466&  0.365& 5.708 \\ 
8&    55.7& 1.631&  11.29&   44.0& 27.3&  0.0068& 0.2133& 0.319&  0.639&  0.315& 5.980 \\ 
9&    42.5& 1.637&  13.57&   75.4& 15.6&  -0.0097& 0.1781& 0.260&  0.243&  0.499& 5.521 \\ 
10&    58.5& 1.747&  12.24&   23.5& 46.5&  0.0050& 0.2108& 0.289&  0.632&  0.290& 5.842 \\ 
11&    62.0& 1.751&  13.37&   21.0& 48.9&  0.0051& 0.1934& 0.271&  0.613&  0.291& 5.809 \\ 
12&    69.0& 1.628&  12.13&   66.0& 18.4&  -0.0037& 0.1982& 0.288&  0.385&  0.403& 5.649 \\ 
13&    55.6& 1.727&  11.95&   74.1& 17.2&  -0.0046& 0.2134& 0.284&  0.262&  0.448& 5.638 \\ 
14&    58.4& 1.557&  11.70&   36.4& 31.3&  0.0059& 0.1965& 0.308&  0.688&  0.305& 5.822 \\ 
15&    65.3& 1.680&  11.06&   69.2& 18.7&  -0.0047& 0.2243& 0.300&  0.337&  0.403& 5.612 \\ 
16&    45.1& 1.674&  12.26&   52.7& 22.9&  0.0087& 0.2016& 0.309&  0.559&  0.345& 6.173 \\ 
17&    56.5& 1.686&  12.29&   72.0& 17.2&  -0.0048& 0.2026& 0.281&  0.295&  0.442& 5.635 \\ 
18&    56.0& 1.667&  11.93&   62.3& 19.8&  0.0006& 0.2064& 0.299&  0.438&  0.379& 5.858 \\ 
19&    67.6& 1.561&  11.65&   47.5& 24.8&  0.0034& 0.1979& 0.309&  0.613&  0.328& 5.820 \\ 
20&    61.6& 1.604&  9.99&   42.3& 30.0&  0.0024& 0.2371& 0.337&  0.640&  0.309& 5.837 \\ 
\midrule
Ave&  59.1 & 1.653&  11.72&   49.3& 28.3& 0.0008  & 0.2093& 0.299&  0.511&  0.353& 5.759 \\ 
Std& 7.0 &   0.062&  0.93&   18.2& 11.9& 0.0053  & 0.0158& 0.019&  0.142&  0.060& 0.183 \\ 
\bottomrule
\end{tabular}
\end{table*}

\begin{table*} 
\centering
\caption{Summary of Best Fits for Model $\theta_{37}$ \label{tab:Jdata}}
\begin{tabular}{rccccccccccc}
\toprule
Model & \chis & $M$ & $R$ & $i$ & $\theta$ & $\phi$ & $M/R$ & $\sin i \sin\theta$ & $\cos i \cos\theta$ & Amp & $\beta$\\
& 59 dof & \Msun & km & deg. & deg. & $\times 10^{-3}$ & & & & & $\times10^{-2}$ \\
\midrule[\heavyrulewidth]
True&&       1.600&  12.00&   60.0& 37.0&  0.000& 0.1969& 0.521&  0.399&  0.720& 10.101 \\ 
\midrule  
1&    58.7& 1.710&  15.29&   26.4& 69.2&  -0.0076& 0.1652& 0.416&  0.318&  0.737& 9.765 \\ 
2&    62.1& 1.577&  11.47&   40.6& 56.8&  0.0019& 0.2030& 0.544&  0.416&  0.718& 10.185 \\ 
3&    52.3& 1.594&  11.88&   43.5& 52.1&  0.0035& 0.1981& 0.542&  0.446&  0.700& 10.430 \\ 
4&    68.8& 1.614&  12.43&   34.4& 62.5&  -0.0013& 0.1918& 0.501&  0.381&  0.724& 9.976 \\ 
5&    55.5& 1.640&  11.78&   62.4& 36.8&  0.0007& 0.2056& 0.530&  0.371&  0.736& 10.237 \\ 
6&    44.1& 1.684&  12.68&   68.1& 31.5&  -0.0035& 0.1961& 0.485&  0.318&  0.758& 9.925 \\ 
7&    62.3& 1.611&  11.87&   36.6& 61.5&  0.0001& 0.2004& 0.524&  0.383&  0.730& 10.103 \\ 
8&    51.2& 1.651&  12.22&   35.8& 62.1&  0.0005& 0.1995& 0.518&  0.379&  0.728& 10.259 \\ 
9&    70.9& 1.516&  11.91&   40.8& 53.4&  0.0000& 0.1880& 0.524&  0.452&  0.696& 9.935 \\ 
10&    65.4& 1.555&  11.89&   58.0& 38.0&  -0.0008& 0.1931& 0.521&  0.418&  0.712& 9.953 \\ 
11&    50.3& 1.531&  12.34&   54.2& 39.2&  0.0004& 0.1832& 0.513&  0.454&  0.690& 9.994 \\ 
12&    70.3& 1.567&  11.64&   56.8& 39.8&  0.0006& 0.1988& 0.536&  0.421&  0.714& 10.102 \\ 
13&    66.2& 1.743&  13.42&   69.9& 29.7&  -0.0046& 0.1918& 0.465&  0.299&  0.763& 9.995 \\ 
14&    63.6& 1.711&  13.02&   67.0& 31.7&  -0.0023& 0.1941& 0.483&  0.333&  0.743& 10.111 \\ 
15&    85.4& 1.836&  13.52&   74.9& 28.2&  -0.0060& 0.2005& 0.456&  0.230&  0.808& 10.026 \\ 
16&    63.9& 1.540&  12.21&   47.7& 45.4&  0.0021& 0.1863& 0.527&  0.472&  0.685& 10.209 \\ 
17&    60.4& 1.526&  12.01&   50.5& 43.1&  0.0016& 0.1876& 0.528&  0.464&  0.690& 10.083 \\ 
18&    48.3& 1.586&  11.67&   46.5& 49.4&  0.0039& 0.2007& 0.551&  0.448&  0.703& 10.448 \\ 
19&    49.5& 1.575&  11.69&   37.2& 60.6&  -0.0010& 0.1990& 0.526&  0.391&  0.729& 9.973 \\ 
20&    73.3& 1.695&  12.81&   31.8& 67.4&  -0.0023& 0.1954& 0.487&  0.327&  0.752& 10.043 \\ 
\midrule
Ave&  61.1 & 1.623&  12.39&   49.1& 47.9& -0.0007  & 0.1939& 0.509&  0.386&  0.726& 10.088 \\ 
Std& 9.9 &   0.082&  0.88&   13.8& 13.0& 0.0029  & 0.0088& 0.033&  0.064&  0.029& 0.165 \\ 
\bottomrule
\end{tabular}
\end{table*}

\begin{table*} 
\centering
\caption{Summary of Best Fits for Model $\theta_{37}$O \label{tab:JOdata}}
\begin{tabular}{rccccccccccc}
\toprule
Model & \chis & $M$ & $R$ & $i$ & $\theta$ & $\phi$ & $M/R$ & $\sin i \sin\theta$ & $\cos i \cos\theta$ & Amp & $\beta$\\
& 59 dof & \Msun & km & deg. & deg. & $\times 10^{-3}$ & & & & & $\times10^{-2}$ \\
\midrule[\heavyrulewidth]
True&&       1.600&  12.00&   60.0& 37.0&  0.000& 0.1969& 0.521&  0.399&  0.720& 10.101 \\ 
\midrule  
1&    59.5& 1.611&  11.86&   46.0& 48.3&  0.0041& 0.2006& 0.537&  0.462&  0.674& 10.351 \\ 
2&    58.7& 1.534&  11.65&   53.7& 41.8&  0.0014& 0.1945& 0.537&  0.441&  0.707& 10.068 \\ 
3&    59.7& 1.614&  11.99&   37.4& 56.7&  -0.0002& 0.1988& 0.507&  0.436&  0.663& 9.858 \\ 
4&    74.1& 1.546&  11.50&   56.3& 40.6&  0.0010& 0.1985& 0.541&  0.422&  0.721& 10.081 \\ 
5&    65.3& 1.617&  11.65&   40.2& 54.5&  0.0015& 0.2050& 0.526&  0.443&  0.665& 10.032 \\ 
6&    51.4& 1.560&  11.29&   61.6& 37.9&  -0.0008& 0.2041& 0.540&  0.375&  0.750& 9.968 \\ 
7&    63.5& 1.561&  11.71&   62.1& 37.0&  -0.0002& 0.1969& 0.532&  0.374&  0.762& 10.064 \\ 
8&    55.8& 1.611&  11.74&   44.2& 50.2&  0.0034& 0.2026& 0.536&  0.459&  0.670& 10.254 \\ 
9&    42.1& 1.520&  11.89&   58.1& 38.6&  0.0004& 0.1888& 0.530&  0.413&  0.740& 10.039 \\ 
10&    56.0& 1.576&  13.48&   33.6& 58.5&  0.0003& 0.1727& 0.472&  0.435&  0.676& 9.893 \\ 
11&    59.5& 1.608&  11.85&   45.8& 48.5&  0.0039& 0.2004& 0.537&  0.462&  0.675& 10.343 \\ 
12&    57.7& 1.562&  11.68&   59.3& 38.7&  0.0010& 0.1975& 0.537&  0.398&  0.741& 10.147 \\ 
13&    59.7& 1.613&  11.98&   37.4& 56.7&  -0.0002& 0.1988& 0.508&  0.436&  0.663& 9.859 \\ 
14&    74.1& 1.544&  11.50&   56.3& 40.6&  0.0009& 0.1983& 0.541&  0.422&  0.721& 10.072 \\ 
15&    65.3& 1.618&  11.64&   40.3& 54.5&  0.0015& 0.2053& 0.526&  0.443&  0.665& 10.034 \\ 
16&    51.4& 1.564&  11.22&   62.0& 37.9&  -0.0009& 0.2059& 0.542&  0.371&  0.753& 9.979 \\ 
17&    63.4& 1.566&  11.64&   61.4& 37.6&  0.0002& 0.1987& 0.535&  0.380&  0.754& 10.094 \\ 
18&    55.7& 1.617&  11.74&   43.5& 51.0&  0.0034& 0.2034& 0.535&  0.457&  0.668& 10.246 \\ 
19&    42.1& 1.521&  11.88&   58.0& 38.7&  0.0005& 0.1891& 0.530&  0.414&  0.738& 10.042 \\ 
20&    63.6& 1.652&  11.89&   38.4& 56.6&  0.0016& 0.2052& 0.518&  0.432&  0.665& 10.089 \\ 
\midrule
Ave&  58.9 & 1.581&  11.79&   49.8& 46.3& 0.0011  & 0.1982& 0.528&  0.424&  0.704& 10.076 \\ 
Std& 8.1 &   0.037&  0.44&   9.7& 7.8& 0.0015  & 0.0075& 0.016&  0.029&  0.037& 0.136 \\ 
\bottomrule
\end{tabular}
\end{table*}

\begin{table*} 
\centering
\caption{Summary of Best Fits for Model $\theta_{60}$ \label{tab:Edata}}
\begin{tabular}{rccccccccccc}
\toprule
Model & \chis & $M$ & $R$ & $i$ & $\theta$ & $\phi$ & $M/R$ & $\sin i \sin\theta$ & $\cos i \cos\theta$ & Amp & $\beta$\\
& 59 dof & \Msun & km & deg. & deg. & $\times 10^{-3}$ & & & & & $\times10^{-2}$ \\
\midrule[\heavyrulewidth]
True&&       1.600&  12.00&   60.0& 60.0&  0.000& 0.1969& 0.750&  0.250&  1.305& 14.536 \\ 
\midrule  
1&    49.6& 1.635&  12.08&   67.7& 53.5&  -0.0005& 0.1999& 0.744&  0.225&  1.332& 14.584 \\ 
2&    62.0& 1.596&  12.10&   59.8& 59.6&  -0.0001& 0.1948& 0.745&  0.255&  1.299& 14.516 \\ 
3&    56.2& 1.621&  12.21&   68.0& 52.1&  -0.0020& 0.1961& 0.732&  0.230&  1.323& 14.411 \\ 
4&    72.0& 1.599&  12.05&   63.3& 56.5&  -0.0004& 0.1960& 0.745&  0.248&  1.307& 14.479 \\ 
5&    67.5& 1.598&  11.89&   63.8& 56.8&  -0.0003& 0.1985& 0.751&  0.242&  1.315& 14.453 \\ 
6&    60.7& 1.593&  12.01&   59.6& 60.0&  -0.0002& 0.1959& 0.747&  0.253&  1.300& 14.472 \\ 
7&    65.9& 1.598&  12.10&   59.7& 59.7&  -0.0001& 0.1950& 0.746&  0.254&  1.299& 14.527 \\ 
8&    53.9& 1.626&  12.12&   66.3& 54.3&  -0.0003& 0.1981& 0.743&  0.235&  1.320& 14.579 \\ 
9&    43.0& 1.587&  12.15&   59.6& 59.3&  -0.0004& 0.1929& 0.741&  0.259&  1.294& 14.449 \\ 
10&    58.8& 1.627&  12.43&   51.1& 68.3&  -0.0014& 0.1933& 0.723&  0.232&  1.320& 14.424 \\ 
11&    63.0& 1.617&  12.12&   61.0& 58.9&  0.0003& 0.1970& 0.749&  0.250&  1.303& 14.673 \\ 
12&    45.1& 1.589&  12.34&   63.8& 54.4&  -0.0010& 0.1902& 0.729&  0.257&  1.293& 14.374 \\ 
13&    50.1& 1.613&  11.97&   56.7& 64.0&  0.0001& 0.1990& 0.751&  0.241&  1.314& 14.567 \\ 
14&    61.7& 1.597&  12.04&   57.3& 62.5&  -0.0005& 0.1959& 0.746&  0.249&  1.306& 14.490 \\ 
15&    60.2& 1.622&  12.08&   65.7& 55.0&  -0.0004& 0.1983& 0.746&  0.236&  1.322& 14.594 \\ 
16&    66.3& 1.607&  12.06&   60.3& 59.7&  0.0002& 0.1968& 0.750&  0.250&  1.304& 14.598 \\ 
17&    43.7& 1.588&  11.99&   59.6& 60.0&  -0.0003& 0.1956& 0.747&  0.253&  1.302& 14.441 \\ 
18&    49.9& 1.606&  12.23&   59.3& 59.8&  0.0002& 0.1939& 0.743&  0.257&  1.295& 14.606 \\ 
19&    55.6& 1.612&  11.84&   63.1& 58.3&  0.0003& 0.2011& 0.759&  0.238&  1.322& 14.610 \\ 
20&    69.1& 1.626&  12.00&   64.0& 57.0&  0.0004& 0.2001& 0.754&  0.239&  1.317& 14.689 \\ 
\midrule
Ave&  57.7 & 1.608&  12.09&   61.5& 58.5& -0.0003  & 0.1964& 0.745&  0.245&  1.309& 14.527 \\ 
Std& 8.4 &   0.014&  0.13&   4.0& 3.7& 0.0006  & 0.0026& 0.008&  0.010&  0.011& 0.087 \\ 
\bottomrule
\end{tabular}
\end{table*}

\begin{table*} 
\centering
\caption{Summary of Best Fits for Model $\theta_{60}C_{6250}$ \label{tab:Ecdata}}
\begin{tabular}{rccccccccccc}
\toprule
Model & \chis & $M$ & $R$ & $i$ & $\theta$ & $\phi$ & $M/R$ & $\sin i \sin\theta$ & $\cos i \cos\theta$ & Amp & $\beta$\\
& 59 dof & \Msun & km & deg. & deg. & $\times 10^{-3}$ & & & & & $\times10^{-2}$ \\
\midrule[\heavyrulewidth]
True&&       1.600&  12.00&   60.0& 60.0&  0.000& 0.1969& 0.750&  0.250&  1.305& 14.536 \\ 
\midrule  
1&    69.3& 1.634&  12.22&   60.8& 59.3&  0.0007& 0.1975& 0.751&  0.249&  1.305& 14.831 \\ 
2&    56.7& 1.625&  12.83&   49.1& 68.7&  -0.0024& 0.1870& 0.704&  0.238&  1.310& 14.351 \\ 
3&    53.8& 1.616&  11.97&   60.7& 60.1&  0.0008& 0.1994& 0.756&  0.244&  1.314& 14.678 \\ 
4&    67.0& 1.615&  12.54&   49.9& 68.6&  -0.0024& 0.1902& 0.712&  0.235&  1.314& 14.269 \\ 
5&    58.0& 1.584&  12.00&   60.5& 59.0&  -0.0005& 0.1949& 0.746&  0.253&  1.302& 14.415 \\ 
6&    45.2& 1.585&  12.87&   52.3& 63.6&  -0.0018& 0.1819& 0.709&  0.272&  1.271& 14.378 \\ 
7&    59.1& 1.577&  13.10&   50.5& 64.2&  -0.0020& 0.1778& 0.695&  0.276&  1.260& 14.270 \\ 
8&    58.3& 1.621&  12.01&   59.1& 61.5&  0.0009& 0.1993& 0.754&  0.245&  1.308& 14.684 \\ 
9&    64.1& 1.615&  12.42&   56.8& 61.6&  0.0006& 0.1920& 0.736&  0.261&  1.286& 14.647 \\ 
10&    59.0& 1.598&  12.49&   63.4& 54.3&  -0.0006& 0.1889& 0.726&  0.261&  1.286& 14.465 \\ 
11&    68.3& 1.605&  12.86&   65.1& 51.4&  -0.0017& 0.1843& 0.709&  0.263&  1.278& 14.430 \\ 
12&    60.6& 1.672&  12.64&   47.8& 72.6&  -0.0021& 0.1953& 0.707&  0.201&  1.356& 14.395 \\ 
13&    52.2& 1.571&  12.50&   58.8& 57.8&  -0.0011& 0.1856& 0.724&  0.276&  1.268& 14.354 \\ 
14&    63.5& 1.632&  12.42&   60.1& 58.9&  0.0012& 0.1940& 0.742&  0.258&  1.292& 14.819 \\ 
15&    51.0& 1.602&  12.16&   59.9& 59.3&  0.0001& 0.1946& 0.744&  0.256&  1.296& 14.558 \\ 
16&    48.5& 1.662&  13.05&   70.4& 47.6&  -0.0021& 0.1881& 0.696&  0.226&  1.319& 14.460 \\ 
17&    61.3& 1.592&  12.30&   55.1& 63.2&  0.0000& 0.1911& 0.733&  0.258&  1.292& 14.415 \\ 
18&    49.8& 1.629&  11.14&   63.8& 62.9&  0.0021& 0.2159& 0.798&  0.201&  1.373& 14.839 \\ 
19&    45.6& 1.697&  13.36&   73.4& 45.0&  -0.0040& 0.1876& 0.678&  0.202&  1.350& 14.410 \\ 
20&    72.0& 1.620&  12.52&   54.9& 63.3&  0.0002& 0.1911& 0.731&  0.258&  1.288& 14.643 \\ 
\midrule
Ave&  58.2 & 1.618&  12.47&   58.6& 60.2& -0.0007  & 0.1918& 0.728&  0.247&  1.303& 14.516 \\ 
Std& 7.7 &   0.031&  0.48&   6.6& 6.6& 0.0015  & 0.0078& 0.027&  0.023&  0.028& 0.179 \\ 
\bottomrule
\end{tabular}
\end{table*}

\begin{table*} 
\centering
\caption{Summary of Best Fits for Model $(M/R)_{hi}$ \label{tab:Kdata}}
\begin{tabular}{rccccccccccc}
\toprule
Model & \chis & $M$ & $R$ & $i$ & $\theta$ & $\phi$ & $M/R$ & $\sin i \sin\theta$ & $\cos i \cos\theta$ & Amp & $\beta$\\
& 59 dof & \Msun & km & deg. & deg. & $\times 10^{-3}$ & & & & & $\times10^{-2}$ \\
\midrule[\heavyrulewidth]
True&&       1.680&  11.00&   60.0& 20.8&  0.000& 0.2255& 0.308&  0.467&  0.350& 5.742 \\ 
\midrule
1&    59.5& 1.670&  11.16&   45.6& 27.2&  0.0075& 0.2210& 0.327&  0.622&  0.321& 6.141 \\ 
2&    58.8& 1.562&  9.98&   55.1& 23.8&  0.0001& 0.2312& 0.331&  0.524&  0.347& 5.657 \\ 
3&    58.4& 1.680&  11.28&   17.0& 70.2&  -0.0099& 0.2199& 0.275&  0.323&  0.384& 5.215 \\ 
4&    72.5& 1.569&  10.34&   29.1& 44.0&  0.0052& 0.2241& 0.338&  0.628&  0.327& 5.918 \\ 
5&    70.9& 1.570&  9.71&   23.5& 57.5&  0.0007& 0.2388& 0.336&  0.492&  0.354& 5.681 \\ 
6&    56.6& 1.484&  8.95&   42.3& 33.5&  0.0038& 0.2448& 0.371&  0.617&  0.338& 5.844 \\ 
7&    64.1& 1.610&  10.36&   24.6& 52.7&  0.0032& 0.2295& 0.331&  0.551&  0.339& 5.855 \\ 
8&    56.2& 1.726&  11.07&   63.8& 19.8&  -0.0004& 0.2303& 0.303&  0.416&  0.360& 5.748 \\ 
9&    41.3& 1.543&  10.91&   36.6& 33.0&  0.0050& 0.2089& 0.325&  0.673&  0.315& 5.848 \\ 
10&    63.7& 1.669&  10.75&   21.8& 58.5&  0.0018& 0.2293& 0.316&  0.485&  0.348& 5.813 \\ 
11&    62.4& 1.949&  13.56&   14.4& 76.1&  -0.0104& 0.2123& 0.241&  0.232&  0.402& 5.427 \\ 
12&    70.5& 1.520&  10.11&   36.5& 35.0&  0.0041& 0.2220& 0.341&  0.659&  0.322& 5.814 \\ 
13&    56.7& 1.510&  8.99&   46.3& 30.9&  0.0051& 0.2482& 0.372&  0.592&  0.343& 5.922 \\ 
14&    61.9& 1.525&  10.31&   41.0& 30.7&  0.0042& 0.2184& 0.335&  0.649&  0.323& 5.784 \\ 
15&    64.7& 1.578&  9.35&   22.7& 62.4&  -0.0004& 0.2492& 0.342&  0.428&  0.370& 5.673 \\ 
16&    45.5& 1.733&  11.33&   48.7& 25.8&  0.0089& 0.2259& 0.327&  0.594&  0.325& 6.287 \\ 
17&    53.3& 1.547&  10.54&   34.9& 35.9&  0.0058& 0.2168& 0.335&  0.665&  0.320& 5.895 \\ 
18&    57.0& 1.570&  9.94&   45.1& 29.4&  0.0050& 0.2333& 0.348&  0.615&  0.331& 5.958 \\ 
19&    69.5& 1.629&  11.12&   23.1& 52.7&  0.0020& 0.2163& 0.312&  0.558&  0.332& 5.785 \\ 
20&    62.0& 1.524&  9.19&   25.7& 54.7&  0.0010& 0.2450& 0.354&  0.520&  0.354& 5.724 \\ 
\midrule
Ave&  60.3 & 1.608&  10.45&   34.9& 42.7& 0.0021  & 0.2282& 0.328&  0.542&  0.343& 5.799 \\ 
Std& 7.7 &   0.105&  1.03&   13.1& 16.2& 0.0048  & 0.0117& 0.029&  0.116&  0.022& 0.220 \\ 
\bottomrule
\end{tabular}
\end{table*}

\begin{table*} 
\centering
\caption{Summary of Best Fits for Model $(M/R)_{lo}$ \label{tab:Ldata}}
\begin{tabular}{rccccccccccc}
\toprule
Model & \chis & $M$ & $R$ & $i$ & $\theta$ & $\phi$ & $M/R$ & $\sin i \sin\theta$ & $\cos i \cos\theta$ & Amp & $\beta$\\
& 59 dof & \Msun & km & deg. & deg. & $\times 10^{-3}$ & & & & & $\times10^{-2}$ \\
\midrule[\heavyrulewidth]
True&&       1.350&  13.00&   60.0& 19.8&  0.000& 0.1534& 0.293&  0.471&  0.423& 5.746 \\ 
\midrule   
1&    72.0& 1.299&  11.73&   53.6& 23.8&  0.0041& 0.1635& 0.325&  0.543&  0.413& 5.837 \\ 
2&    79.5& 1.393&  13.44&   18.2& 63.5&  -0.0017& 0.1531& 0.280&  0.423&  0.435& 5.685 \\ 
3&    55.7& 1.357&  11.28&   55.8& 24.0&  0.0043& 0.1777& 0.336&  0.514&  0.426& 5.945 \\ 
4&    45.9& 1.335&  14.86&   16.6& 63.8&  -0.0040& 0.1327& 0.256&  0.423&  0.424& 5.578 \\ 
5&    60.9& 1.423&  14.18&   17.5& 64.2&  -0.0019& 0.1482& 0.271&  0.415&  0.434& 5.765 \\ 
6&    49.1& 1.191&  12.43&   32.8& 37.5&  0.0089& 0.1415& 0.330&  0.667&  0.382& 6.092 \\ 
7&    73.8& 1.275&  12.72&   56.2& 21.2&  0.0011& 0.1480& 0.300&  0.519&  0.412& 5.725 \\ 
8&    67.7& 1.225&  12.85&   21.5& 54.4&  0.0011& 0.1408& 0.297&  0.542&  0.403& 5.672 \\ 
9&    51.4& 1.127&  12.90&   52.5& 21.6&  0.0004& 0.1290& 0.292&  0.566&  0.395& 5.499 \\ 
10&    71.8& 1.236&  11.96&   52.0& 23.8&  0.0025& 0.1526& 0.318&  0.563&  0.406& 5.739 \\ 
11&    61.0& 1.235&  15.25&   18.6& 56.5&  0.0016& 0.1196& 0.266&  0.523&  0.392& 5.854 \\ 
12&    67.3& 1.115&  13.10&   27.3& 42.4&  0.0069& 0.1257& 0.309&  0.656&  0.375& 5.890 \\ 
13&    81.0& 1.307&  14.36&   19.4& 57.3&  0.0019& 0.1344& 0.279&  0.510&  0.402& 5.890 \\ 
14&    65.0& 1.158&  10.84&   35.0& 37.5&  0.0057& 0.1578& 0.350&  0.649&  0.397& 5.760 \\ 
15&    61.6& 1.272&  13.07&   58.3& 19.8&  -0.0001& 0.1437& 0.288&  0.495&  0.413& 5.606 \\ 
16&    58.2& 1.527&  11.45&   67.0& 20.7&  0.0021& 0.1969& 0.326&  0.366&  0.471& 6.020 \\ 
17&    47.4& 1.311&  14.88&   17.0& 62.7&  -0.0041& 0.1301& 0.259&  0.439&  0.421& 5.638 \\ 
18&    69.6& 1.200&  13.22&   24.9& 47.0&  0.0072& 0.1340& 0.309&  0.618&  0.385& 5.996 \\ 
19&    59.3& 1.417&  14.13&   17.5& 64.3&  -0.0027& 0.1481& 0.271&  0.413&  0.435& 5.744 \\ 
20&    51.7& 1.183&  12.26&   24.6& 49.2&  0.0037& 0.1425& 0.315&  0.594&  0.397& 5.741 \\ 
\midrule
Ave&  62.5 & 1.279&  13.05&   34.3& 42.8& 0.0019  & 0.1460& 0.299&  0.522&  0.411& 5.784 \\ 
Std& 10.1 &   0.105&  1.23&   17.2& 17.1& 0.0036  & 0.0178& 0.027&  0.086&  0.022& 0.152 \\ 
\bottomrule
\end{tabular}
\end{table*}

\begin{table*} 
\centering
\caption{Summary of Best Fits for Model $\beta_{hi}(M/R)_{hi}$ \label{tab:Mdata}}
\begin{tabular}{rccccccccccc}
\toprule
Model & \chis & $M$ & $R$ & $i$ & $\theta$ & $\phi$ & $M/R$ & $\sin i \sin\theta$ & $\cos i \cos\theta$ & Amp & $\beta$\\
& 59 dof & \Msun & km & deg. & deg. & $\times 10^{-3}$ & & & & & $\times10^{-2}$ \\
\midrule[\heavyrulewidth]
True&&       1.680&  11.00&   60.0& 64.0&  0.000& 0.2255& 0.778&  0.219&  1.235& 14.533 \\ 
\midrule  
1&    65.6& 1.758&  11.85&   74.5& 48.5&  -0.0028& 0.2191& 0.722&  0.177&  1.274& 14.351 \\ 
2&    68.2& 1.690&  11.13&   65.6& 58.0&  -0.0001& 0.2242& 0.772&  0.219&  1.234& 14.553 \\ 
3&    44.0& 1.736&  11.41&   51.9& 72.6&  -0.0011& 0.2247& 0.751&  0.185&  1.267& 14.515 \\ 
4&    57.4& 1.676&  11.32&   58.7& 62.8&  -0.0009& 0.2186& 0.760&  0.237&  1.215& 14.428 \\ 
5&    51.9& 1.693&  11.17&   68.2& 55.4&  -0.0011& 0.2238& 0.765&  0.210&  1.242& 14.451 \\ 
6&    64.3& 1.686&  11.21&   67.7& 55.4&  -0.0012& 0.2221& 0.761&  0.216&  1.237& 14.393 \\ 
7&    67.5& 1.680&  10.96&   64.1& 60.2&  0.0000& 0.2264& 0.781&  0.217&  1.239& 14.548 \\ 
8&    54.0& 1.696&  11.22&   54.3& 69.4&  -0.0009& 0.2232& 0.760&  0.206&  1.248& 14.407 \\ 
9&    49.4& 1.687&  11.10&   56.1& 67.8&  -0.0004& 0.2244& 0.768&  0.211&  1.242& 14.443 \\ 
10&    41.0& 1.701&  11.22&   55.3& 68.3&  -0.0008& 0.2239& 0.764&  0.211&  1.240& 14.505 \\ 
11&    33.6& 1.692&  11.01&   67.4& 57.2&  0.0000& 0.2269& 0.777&  0.208&  1.246& 14.548 \\ 
12&    44.0& 1.693&  11.04&   64.2& 60.1&  0.0004& 0.2265& 0.781&  0.217&  1.237& 14.650 \\ 
13&    77.8& 1.655&  11.27&   62.5& 58.6&  -0.0017& 0.2169& 0.757&  0.241&  1.212& 14.247 \\ 
14&    54.0& 1.696&  11.22&   54.3& 69.4&  -0.0009& 0.2232& 0.760&  0.206&  1.248& 14.407 \\ 
15&    47.3& 1.749&  11.90&   72.7& 49.4&  -0.0025& 0.2170& 0.724&  0.194&  1.254& 14.410 \\ 
16&    54.3& 1.700&  11.06&   55.6& 69.2&  -0.0003& 0.2270& 0.772&  0.200&  1.253& 14.522 \\ 
17&    54.2& 1.693&  10.88&   58.2& 67.5&  0.0005& 0.2298& 0.785&  0.202&  1.252& 14.612 \\ 
18&    44.0& 1.714&  11.32&   69.3& 54.3&  -0.0006& 0.2236& 0.759&  0.206&  1.243& 14.540 \\ 
19&    60.3& 1.672&  11.38&   58.9& 62.1&  -0.0009& 0.2170& 0.757&  0.242&  1.211& 14.397 \\ 
20&    43.6& 1.681&  11.14&   58.0& 65.1&  -0.0006& 0.2228& 0.769&  0.223&  1.230& 14.470 \\ 
\midrule
Ave&  53.8 & 1.697&  11.24&   61.9& 61.6& -0.0008  & 0.2231& 0.762&  0.211&  1.241& 14.470 \\ 
Std& 10.7 &   0.024&  0.25&   6.5& 6.8& 0.0008  & 0.0036& 0.016&  0.016&  0.016& 0.093 \\ 
\bottomrule
\end{tabular}
\end{table*}

\begin{table*} 
\centering
\caption{Summary of Best Fits for Model $\nu_{400}$ \label{tab:Udata}}
\begin{tabular}{rccccccccccc}
\toprule
Model & \chis & $M$ & $R$ & $i$ & $\theta$ & $\phi$ & $M/R$ & $\sin i \sin\theta$ & $\cos i \cos\theta$ & Amp & $\beta$\\
& 59 dof & \Msun & km & deg. & deg. & $\times 10^{-3}$ & & & & & $\times10^{-2}$ \\
\midrule[\heavyrulewidth]
True&&       1.600&  12.00&   60.0& 20.0&  0.000& 0.1969& 0.296&  0.470&  0.373& 3.827 \\ 
\midrule   
1&    45.9& 1.082&  15.98&   30.4& 30.5&  0.0053& 0.1000& 0.257&  0.743&  0.296& 3.851 \\ 
2&    67.0& 1.597&  11.03&   19.6& 65.3&  -0.0012& 0.2138& 0.304&  0.394&  0.396& 3.715 \\ 
3&    46.0& 1.527&  11.63&   32.5& 38.3&  0.0068& 0.1939& 0.332&  0.662&  0.339& 4.142 \\ 
4&    75.8& 1.740&  14.38&   66.5& 16.0&  -0.0028& 0.1787& 0.253&  0.384&  0.382& 3.805 \\ 
5&    54.8& 1.475&  9.59&   42.5& 33.6&  0.0052& 0.2271& 0.374&  0.614&  0.363& 4.070 \\ 
6&    45.4& 1.898&  16.00&   76.9& 12.6&  -0.0105& 0.1752& 0.212&  0.222&  0.432& 3.530 \\ 
7&    59.5& 1.905&  14.61&   15.3& 71.8&  -0.0026& 0.1926& 0.251&  0.301&  0.409& 3.921 \\ 
8&    46.5& 1.581&  12.68&   37.6& 31.8&  0.0083& 0.1841& 0.322&  0.673&  0.334& 4.304 \\ 
9&    67.1& 1.441&  11.71&   34.7& 34.6&  0.0048& 0.1817& 0.323&  0.677&  0.336& 3.977 \\ 
10&    61.3& 1.681&  16.00&   64.2& 15.1&  -0.0043& 0.1552& 0.234&  0.420&  0.362& 3.778 \\ 
11&    50.8& 1.242&  11.45&   28.7& 38.8&  0.0000& 0.1602& 0.301&  0.683&  0.327& 3.504 \\ 
12&    50.1& 1.754&  14.74&   14.8& 69.7&  -0.0063& 0.1757& 0.240&  0.336&  0.395& 3.680 \\ 
13&    57.9& 1.343&  16.00&   35.1& 27.8&  0.0085& 0.1240& 0.268&  0.723&  0.302& 4.152 \\ 
14&    67.4& 1.518&  13.79&   19.7& 53.7&  0.0009& 0.1626& 0.272&  0.557&  0.341& 3.827 \\ 
15&    78.2& 1.715&  16.00&   59.0& 17.0&  0.0017& 0.1583& 0.251&  0.493&  0.346& 4.069 \\ 
16&    64.3& 1.312&  8.27&   38.0& 39.4&  0.0011& 0.2342& 0.390&  0.609&  0.372& 3.714 \\ 
17&    58.4& 1.484&  10.98&   22.1& 56.1&  0.0004& 0.1996& 0.312&  0.517&  0.367& 3.701 \\ 
18&    48.5& 1.495&  8.94&   26.8& 57.7&  0.0041& 0.2470& 0.382&  0.477&  0.395& 4.020 \\ 
19&    50.1& 1.754&  14.74&   14.8& 69.7&  -0.0063& 0.1757& 0.240&  0.336&  0.395& 3.680 \\ 
20&    67.3& 1.497&  12.42&   34.1& 34.5&  0.0073& 0.1780& 0.317&  0.683&  0.331& 4.116 \\ 
\midrule
Ave&  58.1 & 1.552&  13.05&   35.7& 40.7& 0.0010  & 0.1809& 0.292&  0.525&  0.361& 3.878 \\ 
Std& 9.9 &   0.207&  2.46&   17.6& 18.8& 0.0052  & 0.0338& 0.051&  0.155&  0.035& 0.216 \\ 
\bottomrule
\end{tabular}
\end{table*}